\documentclass{aa}  

\usepackage{graphicx}
\usepackage{txfonts}
\usepackage{booktabs}
\usepackage{multirow}
\usepackage{array}
\usepackage{hyperref}
\hypersetup{hidelinks}

\begin{document}

\title{The segmented spiral structure of the Solar neighbourhood traced by young clustered populations}
\titlerunning{Segmented spiral structure of the Solar neighbourhood}
\authorrunning{Sánchez-Gil \& Alfaro}

\author{M. Carmen Sánchez-Gil\inst{1}\corrauth{mcarmen.sanchez@uca.es}
\and Emilio J. Alfaro \inst{2}\email{emilio@iaa.es}}

\institute{Universidad de Cádiz, Dpto. Estadística e Investigación Operativa, 11406, Cádiz, Spain
\and Instituto de Astrof\'isica de Andaluc\'ia (CSIC), Glorieta de la Astronom\'ia, S/N, 18008, Granada, Spain}

\date{Received 29 June 2026; accepted 16 August 2026}
\abstract
{The nature of the Milky Way spiral pattern remains poorly constrained. Although logarithmic arms provide useful reference curves, the youngest tracers in the solar neighbourhood do not always follow a few smooth, continuous arm loci. Young clusters and star-forming groups also reveal local concentrations, branches, and inter-arm features.}
{We analyse young open clusters in the solar neighbourhood, both independently and combined with young stellar object (YSO)-based groups. We aim to determine whether these young clustered tracers define continuous spiral-arm ridges or instead form shorter structures with partial connections between them.}
{We work in the $(\theta_G,\ln R_G)$ plane, where logarithmic spirals are approximately linear, and identify local overdensities using a density-supported Bayesian Gaussian Mixture Model (BGMM). Published tracks for the Perseus, Local, Sagittarius, and Scutum arms are introduced only afterwards as reference curves. We then apply a Minimum Spanning Tree (MST) analysis in the heliocentric $(X,Y)$ plane to examine how these tracers connect in physical space.}
{The open-cluster sample already resolves into several local components. Adding YSO-based groups provides a denser view of recent clustered star formation, highlighting branches and intermediate regions without producing continuous arms. In the MST analysis, the tracers form distinct local branches at small pruning scales, which progressively merge into larger structures as the linking scale increases. The Local--Sagittarius region shows the clearest agreement between both diagnostics: the BGMM identifies an intermediate component between the reference arms, while the MST connects neighbouring branches through this same region.}
{Young structures near the Sun appear as fragmented, spiral-like segments with partial links between neighbouring pieces, rather than as smooth continuations of a few grand-design arms. While we establish this morphology, determining whether it reflects spiral-arm formation mechanisms or subsequent evolution will require additional age, vertical, and kinematic information.}

\maketitle
\nolinenumbers

\section{Introduction}

Understanding the spiral structure of the Milky Way remains a central problem in Galactic astronomy. A fundamental limitation arises from our position inside the Galactic disc: the spiral pattern must be reconstructed from an internal viewpoint, where extinction, crowding, and the non-uniform distribution of tracers strongly affect our ability to identify low-contrast or diffuse structures. Since the first maps based on young stellar populations, classical tracers of recent star formation -- OB stars, high-mass star-forming regions, giant molecular clouds, and young open clusters -- have provided the observational basis for Galactic spiral-arm maps \citep[e.g.][]{Morgan1952, Morgan1953, VdHulst1954, Becker1956, Davies1960, Roberts1969, Roberts1970, Crampton1975, Scoville1975, Georgelin1976, Cohen1980}. These tracers have traditionally been summarised in terms of a few smooth logarithmic arms \citep{Vallee2014, Vallee2017, Xu2018}, closely related to the classical quasi-stationary density-wave picture, in which spiral arms are described as coherent, long-lived structures with a global geometrical form \citep{Lin1964, Lin1966}.

Radio astrometry has been especially important in challenging this traditional view. Trigonometric parallaxes of masers from the BeSSeL Survey and the VERA project have shown that the Local Arm is not merely a minor spur, but a spatially extended structure with a significant pitch angle and level of star-formation activity \citep{Xu2013,Xu2016}. Studies of young tracers have also revealed local high-pitch-angle structures that are difficult to reconcile with a small number of smooth logarithmic arms. A clear example is the kiloparsec-scale filament emerging from the Sagittarius arm reported by \citet{Kuhn2021}. Other young-tracer studies have similarly emphasised the presence of local branches and arm substructure in the solar neighbourhood \citep[e.g.][]{ Pantaleoni2021, Lemasle2022}.

Three-dimensional studies of the interstellar medium provide a complementary view. Dust and gas maps show that the local environment contains elongated, fragmented, and sometimes nearly parallel structures that do not always coincide with the classical logarithmic arms \citep{Kormann2026, Soler2025, Shen2025, Viscasilla2025}. On larger scales, maps of the perturbed H{\sc I} surface density reveal a non-axisymmetric, multi-armed disc with strong local variations, closer to a ragged or flocculent morphology than to an ideal symmetric grand-design pattern \citep{Levine2006}. Together, these results suggest that the local spiral pattern may be better described by a collection of connected and disconnected star-forming structures rather than by a small set of continuous arm ridges.

The dynamical origin of this morphology remains an open question. Recent hydrodynamic simulations show that spiral arms can behave as transient material features that evolve, fragment, and reconnect \citep{Pettitt2015}. In the stellar disc, recent Gaia-based and spectroscopic studies reveal large-scale perturbations, phase-space substructure, and vertical disturbances in the solar neighbourhood \citep[e.g.][and references therein]{Martinez2022, Alfaro2022, Alfaro2025, Zari2025, Shen2025}. Together with the absence of simple age gradients in young open-cluster populations \citep{Liu2025}, these results favour a picture in which local spiral structure may emerge from evolving and partially connected star-forming segments rather than from rigid, long-lived arms.

Spiral galaxies are commonly classified into flocculent, grand-design, and multi-arm or mixed structures according to the coherence and definition of their arm patterns \citep{Elmegreen1987, Dobbs2014}. Flocculent galaxies exhibit patchy, short, and fragmented arms \citep{Elmegreen1987, Dobbs2008}, whereas grand-design spirals possess two prominent and well-defined main arms \citep{Lin1964}. Mixed-type structures display characteristics of both forms simultaneously, often revealing grand-design features in the near-infrared whilst appearing more flocculent in optical wavebands \citep{Block1994}. Each of these morphological patterns is directly linked to the physical mechanisms that generate the spiral structure \citep{Dobbs2014}, although morphology alone does not uniquely determine their dynamical origin. Local instabilities and sheared perturbations typically give rise to flocculent galaxies \citep{Gerola1978, Dobbs2008}, whereas grand-design spirals are commonly associated with tidal interactions with neighbouring companions, central bar driving, or steady-state density wave theory \citep{Lin1964, Toomre1972}. Furthermore, processes such as swing amplification and stochastic star formation play a crucial role in driving the dynamical evolution and transient behaviour of these spiral configurations \citep{Toomre1981, Dobbs2018}.

In this context, this paper adopts a local and morphological perspective. We do not aim to derive a new global parametrisation of the Milky Way spiral arms. Instead, we investigate whether young clustered tracers naturally define smooth grand-design ridges or whether they separate into local overdensities, branches, bridges, and inter-arm features when they are not forced a priori to follow a small number of continuous logarithmic arms. Published logarithmic arms are therefore used only as large-scale geometrical references against which local departures can be measured, rather than as structures imposed on the data.

To address this question, we have developed a pipeline based on a density-supported Bayesian Gaussian Mixture Model \citep[BGMM; e.g.,][]{Pedregosa2011} that identifies local overdensities in the $(\theta_\mathrm{G}, \ln R_\mathrm{G})$ plane, whilst a Minimum Spanning Tree algorithm \citep [MST; e.g.,][]{Kruskal1956} independently tests their spatial connectivity in the heliocentric $(X,Y)$ plane.

The paper is organised as follows. Section~\ref{sec:data_methods} describes the tracer samples and coordinate systems. Section~\ref{sec:methodology} introduces the BGMM segmentation, the reference-guided interpretation, and the MST connectivity analysis. Section~\ref{sec:results} presents the resulting local spiral morphology. Section~\ref{sec:discussion} discusses the interpretation of the detected structures, and Section~\ref{sec:conclusions} summarises the main conclusions.

\section{Data and coordinate systems}
\label{sec:data_methods}

\begin{table}[t]
\centering
\caption{Working tracer samples used in the analysis.}
\label{tab:sample_definition}
\begin{tabular}{llc}
\toprule
Sample & Content & $N$ \\
\midrule
CG & Cantat-Gaudin-based OC catalogue & 81 \\
Dias & Dias OC catalogue & 171 \\
Hao & Hao OC catalogue & 478 \\
\midrule
OC & CG + Dias + Hao & 730 \\
Kuhn YSO & YSO-based star-forming groups & 390 \\
OC+YSO & OC + Kuhn YSO groups & 1120 \\
\bottomrule
\end{tabular}
\tablefoot{
Open-cluster entries are selected with $\log t \leq 7.0$ and
processed with a vertical $3\sigma$ clipping in $Z$. The numbers
correspond to catalogue entries in the working samples and not
necessarily to unique physical clusters after cross-identification.
}
\end{table}

\subsection{Young clustered tracer samples}\label{sec:samples}

The analysis is based on young clustered tracers of recent star formation in the solar neighbourhood. Owing to their short lifetimes compared with the timescales associated with secular radial migration and large-scale orbital diffusion, young open clusters are expected to remain, in a statistical sense, close to the star-forming structures from which they formed.
Rather than constructing a synthetic mean catalogue, we retain the published entries from each catalogue in order to preserve their catalogue-specific selections and parameter determinations.

The open-cluster sample is then constructed from three Gaia-based catalogues: the Cantat-Gaudin compilation \citep{Cantat2020}, the Dias catalogue \citep{Dias2021}, and the Hao catalogue \citep{Hao2022}. These catalogues are not treated as independent complete censuses, since they differ in their detection strategies, membership assignments, age estimates, distance determinations, and completeness. Instead, they are considered complementary samplings of the local young-cluster population.

The same preprocessing is applied to the three catalogues. We retain entries with $\log t \leq 7.0$, corresponding to ages $\leq 10\,{\rm Myr}$, and remove extreme vertical outliers through a $3\sigma$ clipping in $Z$. All tracers are transformed into a common heliocentric and Galactocentric reference frame, adopting $R_0=8.2\,{\rm kpc}$ \citep{Bland2016} for the Sun--Galactic-centre distance.

After the vertical clipping, the retained samples remain concentrated around the Galactic mid-plane. For completeness, we report the vertical distribution as $Z_{\rm med}=-14.2\,{\rm pc}$ and $\sigma_Z=90.8\,{\rm pc}$ for OC, and $Z_{\rm med}=-7.8\,{\rm pc}$ and $\sigma_Z=76.7\,{\rm pc}$ for OC+YSO.

The resulting open-cluster sample, hereafter OC, contains 730 catalogue entries: 81 from Cantat-Gaudin, 171 from Dias, and 478 from Hao. The working sample is kept at catalogue-entry level rather than collapsed into a fully cross-identified list of unique physical clusters. The quoted numbers therefore refer to catalogue entries and should not be interpreted as a deduplicated cluster census. We retain the published catalogue entries to preserve the original catalogue-specific selections and parameter determinations, and to avoid introducing an additional cross-identification criterion. The same physical cluster may therefore contribute more than once in regions of catalogue overlap. Accordingly, the absolute component weights and local tracer densities are not interpreted as a census of unique clusters, but only as descriptors of the morphology of the combined young-cluster distribution.

The combined sample, labelled OC+YSO, is obtained by adding the 390 YSO-based star-forming groups from \citet{Kuhn2021} to the OC sample. These groups are assigned $\log t=7.0$ in the working table to maintain a consistent age-selected format; this value is not interpreted as an individual age estimate. The final OC+YSO sample contains 1120 tracer entries and provides a denser sampling of recent clustered star formation, particularly along local branches and between the reference spiral arms.

Throughout the paper, OC refers to the open-cluster-only sample, whereas OC+YSO denotes the main combined tracer sample. Their comparison provides a consistency test: structures already present in OC and enhanced by the addition of YSO groups are less likely to arise solely from the YSO selection.

Only the OC and OC+YSO samples are used for the BGMM segmentation, the reference-guided fits, and the MST connectivity analysis. Other structures from the literature, including the Radcliffe Wave and published spur candidates, are displayed only as external references and are not used to impose segmentation, arm assignment, or graph connectivity.

No additional heliocentric-distance cut is imposed. In the final working samples, 95\% of the OC entries lie within $4.70\,{\rm kpc}$ of the Sun and 95\% of the OC+YSO entries within $6.26\,{\rm kpc}$; the most distant objects are located at $7.86$ and $9.58\,{\rm kpc}$, respectively. Thus, throughout this work, the term ``solar neighbourhood'' refers operationally to the local Galactic volume sampled by these catalogues, which can be defined as a circle in the Galactic plane of about $4.5\,{\rm kpc}$ in radius. This volume contains more than 90\% of the overall sample and encompasses all the detected overdensities.

\begin{figure}[t]
    \centering
    \includegraphics[width=\columnwidth]{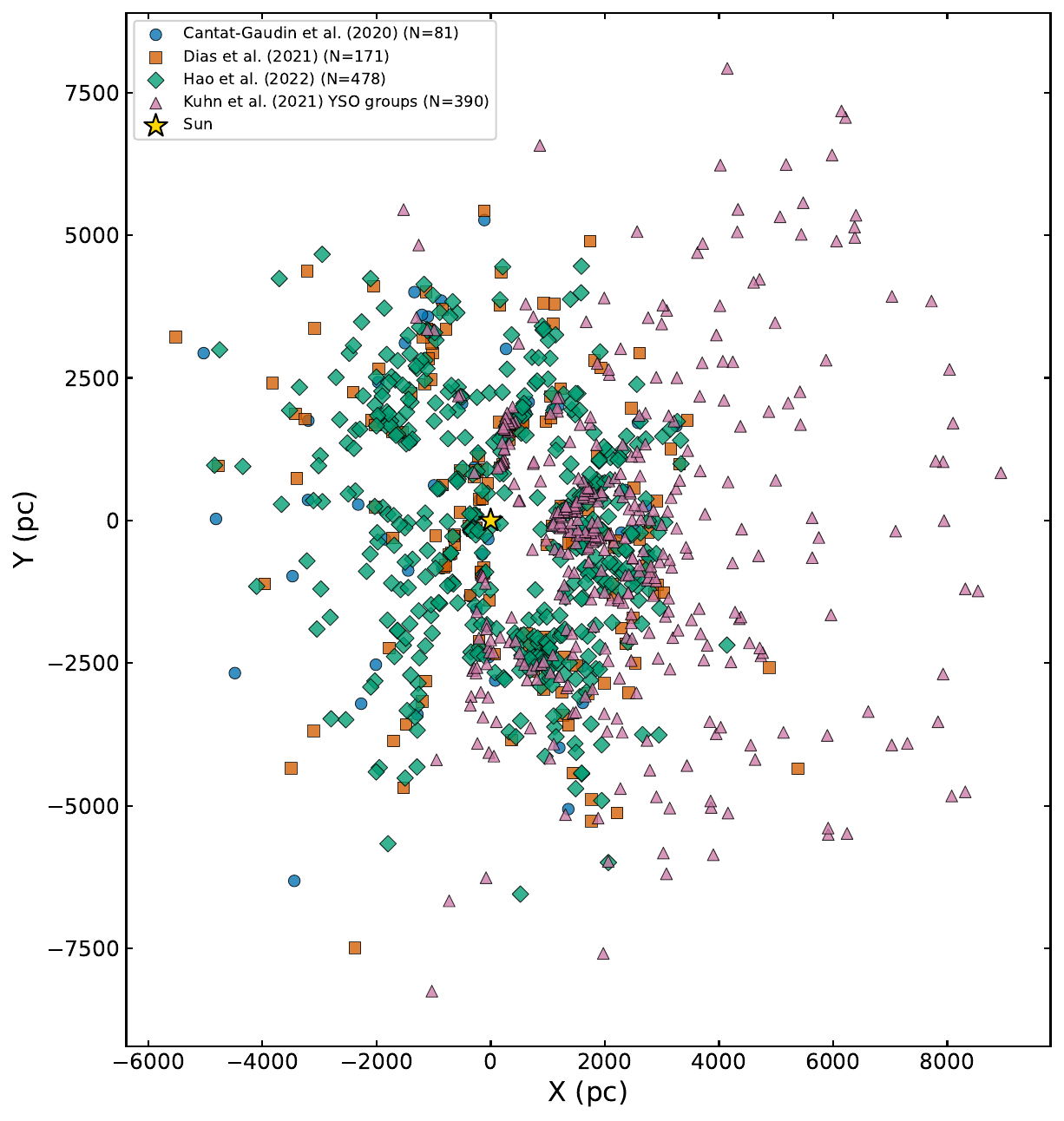}
    \caption{Spatial distribution in the heliocentric $(X,Y)$ plane of all tracer entries listed in Table~\ref{tab:sample_definition}. Different colours and symbols identify the three open-cluster catalogues and the \cite{Kuhn2021} YSO-based groups. The yellow star marks the Sun.}
    \label{fig:input_catalogues_xy}
\end{figure}

Figure~\ref{fig:input_catalogues_xy} shows the heliocentric spatial distribution of the four catalogue contributions listed in Table~\ref{tab:sample_definition}, using distinct symbols and colours to make their different spatial coverage explicit.

\subsection{Observational limitations and scope of the sample}

The reconstruction of the Galactic spiral structure is affected by several observational limitations. Optical Gaia-based catalogues provide an extensive view of nearby young clusters, but remain affected by extinction, crowding, distance-dependent completeness, and the difficulty of identifying embedded or partially disrupted systems. Conversely, YSO-based groups trace very recent star formation but are also subject to their own selection function and do not provide a homogeneous census of the Galactic disc.

The present sample is therefore not intended as an unbiased census of all young Galactic tracers. Instead, it is used to investigate whether the observed local clustered population is more consistent with a small number of smooth logarithmic ridges or with a collection of local overdensities and connections. 
Although the combined effects of extinction and distance might initially suggest that the spatial distribution of open clusters is significantly biased, recent studies demonstrate that young, massive populations ($\log(t/\text{yr}) \le 7.5$) within $4.5\text{ kpc}$ of the Sun suffer minimal observational selection effects. Empirical selection functions for Gaia indicate high completeness for these early-type populations, allowing them to faithfully trace local spiral structure \citep{Cantat2020, Hunt2023, Hunt2025}. This high degree of completeness arises because bright upper main-sequence stars easily exceed the Gaia detection limit; for instance, at a distance modulus $(m - M)_0 \approx 13.27\text{ mag}$ ($d = 4.5\text{ kpc}$), even under a severe extinction of $A_V = 6\text{ mag}$ ($A_G \approx 5.0\text{ mag}$), B5V stars ($M_G \approx -1.0\text{ mag}$) yield $G \approx 17.3\text{ mag}$, remaining comfortably above the Gaia magnitude limit ($G \approx 20\text{ mag}$), while late-type B stars (e.g. B9V, $M_G \approx +0.3\text{ mag}$) reach $G \approx 18.6\text{ mag}$ \citep{CastroGinard2020, Hunt2025}. Injection--recovery tests using algorithms such as HDBSCAN confirm high recovery fractions for young clusters, with age-related completeness drops affecting predominantly older or less massive systems ($\log(t/\text{yr}) > 8.0$) due to stellar fading and dynamical dissolution \citep{Hunt2024, Hunt2025, Cavallo2024}. Moreover, because line-of-sight extinction within $4.5\text{ kpc}$ remains moderate ($A_V < 6\text{ mag}$ outside deep mid-plane sightlines), local dust does not systematically obscure clusters with $M \ge 100\text{--}250\text{ M}_\odot$, leaving large-scale Galactic filaments well preserved \citep{Hunt2023, Hunt2025}.

The density-based and graph-based methods are not designed to define a unique spiral-arm model. Rather, they quantify how young clustered tracers organise themselves when they are not forced a priori to follow continuous grand-design arms. The resulting structures are interpreted as local morphological segments, which may correspond to portions of classical arms or to spurs, bridges, bifurcations, and inter-arm concentrations.

\subsection{Coordinate systems}

The analysis is performed in two complementary coordinate spaces. The heliocentric Cartesian plane $(X,Y)$ is used to visualise the local distribution of tracers and to analyse their connectivity with graph-based methods. The Galactocentric spiral plane $(\theta_G,\ln R_G)$ is used for density segmentation and comparison with logarithmic spiral models.

The transformation to Galactocentric polar coordinates is defined as

\begin{equation}
\theta_G = \arctan2\left(Y,R_0-X\right),
\qquad
R_G = \left[(R_0-X)^2+Y^2\right]^{1/2},
\label{eq:theta_rg_definition}
\end{equation}

where $R_0=8.2\,{\rm kpc}$ and $R_0$, $X$, and $Y$ are expressed in the same distance units. The mixture modelling is performed in the coordinate $\ln R_G$. Since the analysis relies on relative positions and slopes in the $(\theta_G,\ln R_G)$ plane, the choice of distance unit only introduces an additive constant in $\ln R_G$ and does not affect the inferred local slopes.

We adopt the logarithmic-spiral parametrisation of \citet{Castro-Ginard2021},
\begin{equation}
\ln\left(\frac{R_G}{R_{G,\rm ref}}\right)
=
-\left(\theta_G-\theta_{G,\rm ref}\right)\tan\psi ,
\label{eq:logspiral_cg}
\end{equation}
where $R_{G,\rm ref}$ is the Galactocentric radius at the reference azimuth $\theta_{G,\rm ref}$, and $\psi$ is the pitch angle. In the $(\theta_G,\ln R_G)$ plane this relation reduces to a linear form,
\begin{eqnarray}
\ln R_G &=& a + b\theta_G,
\label{eq:log_spiral_linear}
\\
b &=& -\tan\psi .
\label{eq:slope_pitch}
\end{eqnarray}

The linear form (Eq.~\ref{eq:log_spiral_linear}) is used for all numerical fits. The fitted parameters are then expressed in the physical spiral form of Eq.~\ref{eq:logspiral_cg} via
\begin{equation}
R_{G,\rm ref} = \exp\left(a + b\theta_{G,\rm ref}\right).
\label{eq:fit_to_cg_parameters}
\end{equation}
For each reference-guided fit, $\theta_{G,\rm ref}$ is defined as the median Galactocentric azimuth of the tracers entering the fit. It is therefore a reporting anchor rather than a free parameter. All fits in the $(\theta_G,\ln R_G)$ plane are performed with $\theta_G$ in radians; angular values are converted to degrees only for presentation in figures and tables. With this convention, the relation $b=-\tan\psi$ is applied directly without additional rescaling.

Throughout the paper, we use the term segment to denote a local overdensity in the $(\theta_G,\ln R_G)$ plane and/or its projected counterpart in $(X,Y)$. A segment may correspond to a portion of a classical spiral arm, but may also trace a spur, bridge, bifurcation, or inter-arm enhancement. This definition is intentionally local and morphological.

\begin{figure}[!t]
    \centering
    \includegraphics[width=\columnwidth]{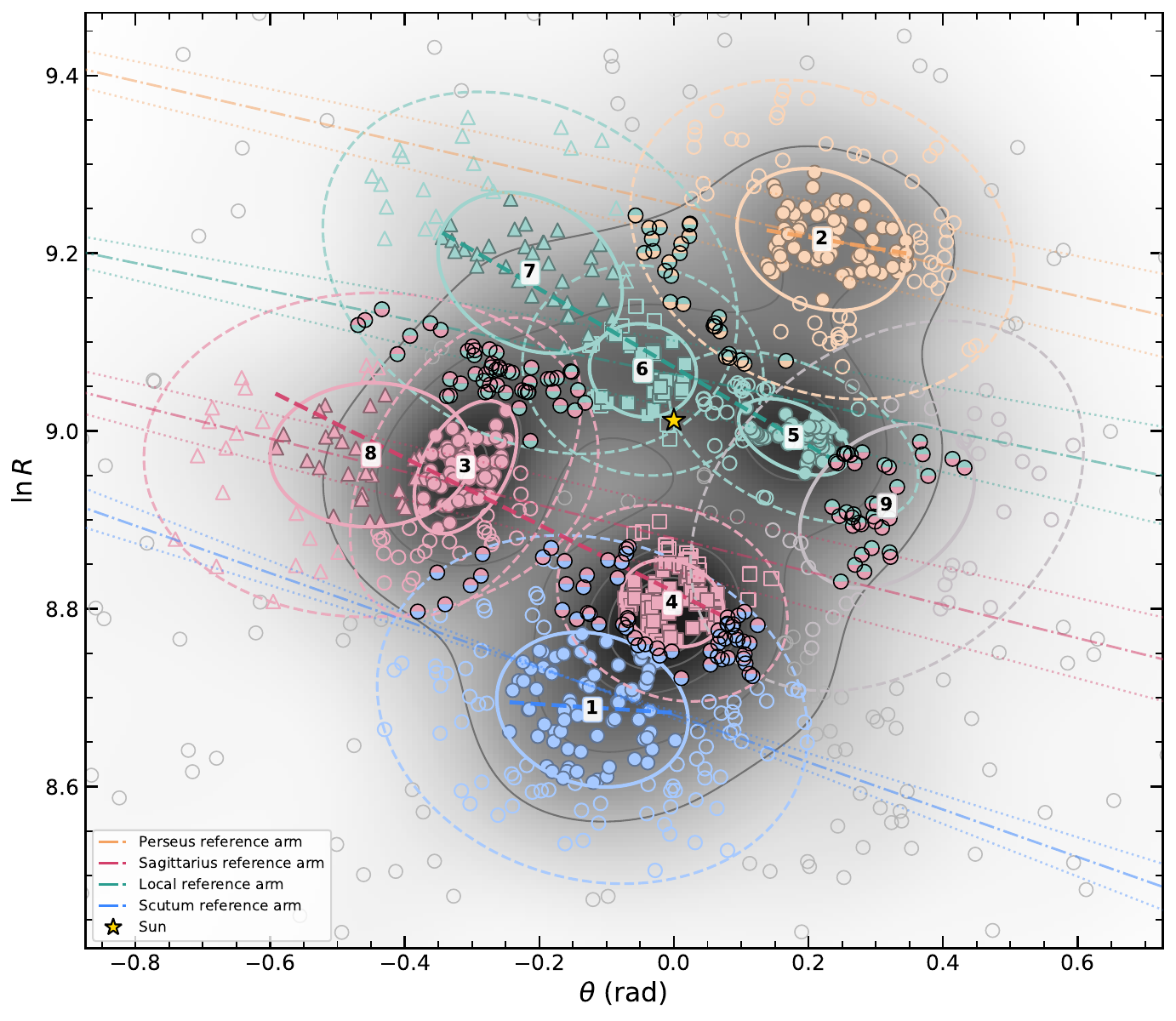}
    \caption{Density segmentation of the fiducial OC+YSO sample in the $(\theta_G,\ln R_G)$ plane. The grey-scale background traces the KDE density field; coloured symbols and ellipses show the retained BGMM components and their membership envelopes. The published spiral-arm loci are shown only for comparison after the segmentation, and the yellow star marks the Sun.}
    \label{fig:theta_lnr_overdensities}
\end{figure}

Figure~\ref{fig:theta_lnr_overdensities} shows the fiducial OC+YSO sample directly in the $(\theta_G,\ln R_G)$ plane used for the segmentation. The grey-scale background represents the smoothed KDE density field, while the coloured BGMM components and their core/envelope regions identify the retained local overdensities. The published spiral-arm loci are overplotted only as a posteriori geometrical references and therefore do not determine which overdensities are selected.

\section{Methods}
\label{sec:methodology}

\begin{table*}[t]
\centering
\caption{Comparison of reference-arm parameters and local fits for the OC and OC+YSO samples.}
\label{tab:reference_guided_fits}
\begin{tabular}{llcccccc}
\toprule
Arm & Sample/source & $N_{\rm fit}$
& $\theta_{G,\rm ref}$ [deg]
& $\theta_{G,\rm range}$ [deg]
& $R_{G,\rm ref}$ [kpc]
& $\psi$ [deg]
& {$R^2_{\rm fit}$} \\
\midrule
\multirow{3}{*}{Perseus}
& Castro-Ginard et al. & 80
& $-13.0$ & $(-20.9,88.2)$
& $10.88\pm0.38$ & $9.8\pm0.9$ & -- \\
& OC & 60
& $12.1$ & $(7.9,18.9)$
& $10.09\pm0.04$ & $5.8\pm3.9$ & $0.04$ \\
& OC+YSO & 70
& $12.9$ & $(7.9,19.8)$
& $10.05\pm0.03$ & $7.2\pm3.0$ & $0.07$ \\
\midrule
\multirow{3}{*}{Sagittarius}
& Castro-Ginard et al. & 79
& $3.5$ & $(-39.3,67.7)$
& $7.10\pm0.26$ & $10.6\pm0.8$ & -- \\
& OC & 96
& $-16.3$ & $(-21.3,7.1)$
& $7.63\pm0.02$ & $23.8\pm0.7$ & $0.90$ \\
& OC+YSO & 176
& $-16.5$ & $(-33.9,4.0)$
& $7.54\pm0.02$ & $20.6\pm0.8$ & $0.78$ \\
\midrule
\multirow{3}{*}{Local}
& Castro-Ginard et al. & 146
& $-2.3$ & $(-26.9,26.6)$
& $8.69\pm0.33$ & $8.9\pm1.3$ & -- \\
& OC & 119
& $-0.7$ & $(-15.7,18.3)$
& $8.78\pm0.02$ & $24.1\pm0.8$ & $0.87$ \\
& OC+YSO & 147
& $0.7$ & $(-19.7,14.3)$
& $8.66\pm0.02$ & $23.7\pm0.7$ & $0.88$ \\
\midrule
\multirow{3}{*}{Scutum}
& Castro-Ginard et al. & 43
& $-4.8$ & $(-32.7,100.9)$
& $6.02\pm0.02$ & $14.9\pm1.6$ & -- \\
& OC & 50
& $-4.8$ & $(-11.7,-1.2)$
& $6.30\pm0.04$ & $-15.1\pm6.4$ & $0.10$ \\
& OC+YSO & 73
& $-7.8$ & $(-14.0,0.1)$
& $5.94\pm0.03$ & $2.8\pm4.8$ & $0.00$ \\
\bottomrule
\end{tabular}
\tablefoot{
The reference-arm parameters are from \citet{Castro-Ginard2021}.
The fitted values are local summaries of the BGMM components
associated a posteriori with each reference arm and are not
intended as a global re-determination of the Galactic spiral
pattern. $N_{\rm fit}$ gives the number of tracers entering each
fit. For the fits obtained in this work, these are the tracers
belonging to the adopted core regions; for the Castro-Ginard
et al. rows, the values are those reported for the corresponding
reference fits. For our fits, the quoted uncertainties are formal
errors from the weighted linear regressions and do not include
systematic effects related to catalogue selection or component
assignment.
}
\end{table*}

All structural units analysed in this work (hereafter segments) correspond to selected BGMM components in the $(\theta_G,\ln R_G)$ space. The analysis pipeline combines density estimation, mixture modelling, geometric characterisation, and independent graph-based connectivity diagnostics.

\subsection{Density-supported BGMM segmentation}

We identify local overdensities in the $(\theta_G,\ln R_G)$ plane using a two-step density–mixture approach. The input catalogues are first subjected to the age selection and to the rejection of extreme vertical outliers in $Z$. Before the KDE and BGMM modelling, Stage~1 applies an additional robust filter in the $(\theta_G,\ln R_G)$ plane to limit the influence of the sparsely sampled extreme tails on the global density and mixture-covariance estimates. The filter is defined independently in each coordinate using three robust standard deviations estimated from the median absolute deviation. In the classification figures, all finite tracers falling within the displayed field are shown, whereas only the objects retained by Stage~1 and lying within the adopted component envelopes receive the corresponding classification colours. First, a kernel density estimator (KDE) is computed to characterise the global density field and to inform the expected complexity of the distribution. The KDE is used for diagnostic purposes only and does not define the segmentation.

We then fit a Bayesian Gaussian mixture model (BGMM) to the standardised $(\theta_G,\ln R_G)$ coordinates. The number of components is guided by the KDE structure and standard information criteria (BIC, AIC). The BGMM provides the actual segmentation of the tracer distribution into localised density-enhanced components (segments).

Each component is classified using mixture weight, effective population, and density support. Components with negligible support are removed, except when they coincide with significant KDE peaks, in which case they are retained as valid density-supported segments. This yields the final set of segments analysed in the rest of the paper.

For each segment $k$, we compute centroid, covariance matrix, angular extent, tracer count, effective membership, and mixture weight. A local orientation is obtained via a weighted linear fit,
\begin{equation}
\ln R_G = a_{{\rm loc},k} + b_{{\rm loc},k}\theta_G .
\label{eq:local_segment_fit_linear}
\end{equation}

The slope $b_{{\rm loc},k}$ is used purely as a geometric descriptor. When needed, it is expressed as a local pitch-like angle $\psi_{{\rm loc},k}=\arctan(-b_{{\rm loc},k})$, without implying global spiral coherence.

\subsection{Segment geometry and membership definition}

For each BGMM component, we define core and envelope regions using the Mahalanobis distance,
\begin{equation}
D_k^2(\mathbf{z}) = (\mathbf{z}-\boldsymbol{\mu}_k)^{\rm T}\boldsymbol{\Sigma}_k^{-1}(\mathbf{z}-\boldsymbol{\mu}_k),
\label{eq:mahalanobis}
\end{equation}
where $\mathbf{z}=(\theta_G,\ln R_G)$, and $\boldsymbol{\mu}_k$ and $\boldsymbol{\Sigma}_k$ are the mean vector and covariance matrix of component $k$.

We define two membership levels: a core region $P_1$ and an extended envelope $P_2$, corresponding to $D_k \leq \sigma_{P_1}$ and $D_k \leq \sigma_{P_2}$, with $\sigma_{P_1}=1.0$ and $\sigma_{P_2}=2.25$. These regions define the membership used for subsequent component characterisation and reference-guided fits. The MST analysis described in Sect. 3.4 is performed independently on the finite tracer positions and does not use these membership assignments.
The $P_1$ region represents the most reliable members of a segment, while $P_2$ traces its extended structure and overlaps with neighbouring segments.

For a two-dimensional Gaussian component, these Mahalanobis radii enclose approximately $39\%$ and $92\%$ of the probability, respectively. We therefore use $P_1$ as a deliberately conservative core and $P_2$ as a broad morphological envelope. These values were chosen as a baseline compromise between isolating the high-density centre of each component and retaining its extended structure; the sensitivity of the results to changes in $\sigma_{P_2}$ is explicitly tested in Appendix~\ref{app:robustness}.

\subsection{Reference-guided interpretation and spiral fits}

Segments are compared a posteriori with the logarithmic spiral arms of \citet{Castro-Ginard2021} (Perseus, Local, Sagittarius, and Scutum). This comparison is purely interpretative and does not influence the segmentation.

Each segment is assigned a provisional association based on proximity to the reference arms in the $(\theta_G,\ln R_G)$ plane. Segments lying between arms or compatible with multiple arms are flagged as inter-arm or bridge candidates. This classification is purely geometrical and does not imply a physical classification scheme.

For each reference arm, we compute a local fit using tracers associated with its assigned segments. The fit is performed in linear form (Eq.~\ref{eq:log_spiral_linear}), and parameters are transformed to the spiral representation of \citet{Castro-Ginard2021} using Eq.~\ref{eq:fit_to_cg_parameters}. Uncertainties on $a$ and $b$ are propagated to $\psi_{\rm fit}$ and $R_{G,\rm ref,fit}$ using the covariance matrix of the linear regression. The reference azimuth $\theta_{G,\rm ref,fit}$ is treated as a fixed reporting quantity and not as a fitted parameter.

These fits quantify local deviations of the segment populations associated with each reference arm with respect to the adopted logarithmic-spiral scaffold.

\subsection{Minimum spanning tree connectivity}

We construct a minimum spanning tree (MST) in the heliocentric $(X,Y)$ plane to quantify spatial connectivity independently of the density segmentation. The MST is fully independent of the BGMM segmentation and does not use segment assignments.

Tracers are treated as nodes in a spatial graph, and the MST provides the unique acyclic network that connects all points with the minimum total edge length. We prune the tree by removing edges longer than a threshold $d_{\rm cut}$, which defines a connectivity scale and is not interpreted as a physical arm width.

We analyse two values of $d_{\rm cut}$, $d_{\rm cut}=200\,{\rm pc}$ and $d_{\rm cut}=300\,{\rm pc}$, to probe the transition from small-scale fragmentation to larger and more connected structures. After pruning, we retain connected components with at least $N_{\rm min}=7$ nodes, suppressing stochastic small groupings while preserving extended branches.

\subsection{Baseline analysis configuration}
\label{sec:baseline_configuration}

The analysis uses a single baseline configuration applied identically to the OC and OC+YSO samples. The parameters are chosen to balance sensitivity to local structure with robustness against over-fragmentation and are not globally optimised.

The BGMM is fitted to the standardised $(\theta_G,\ln R_G)$ coordinates with $\theta_G$ in radians. Segments are defined as the retained mixture components after applying the selection criteria described above. Core and envelope regions are defined using $\sigma_{P_1}=1.0$ and $\sigma_{P_2}=2.25$.

\begin{figure*}[t]
    \centering
    \includegraphics[width=0.95\textwidth]{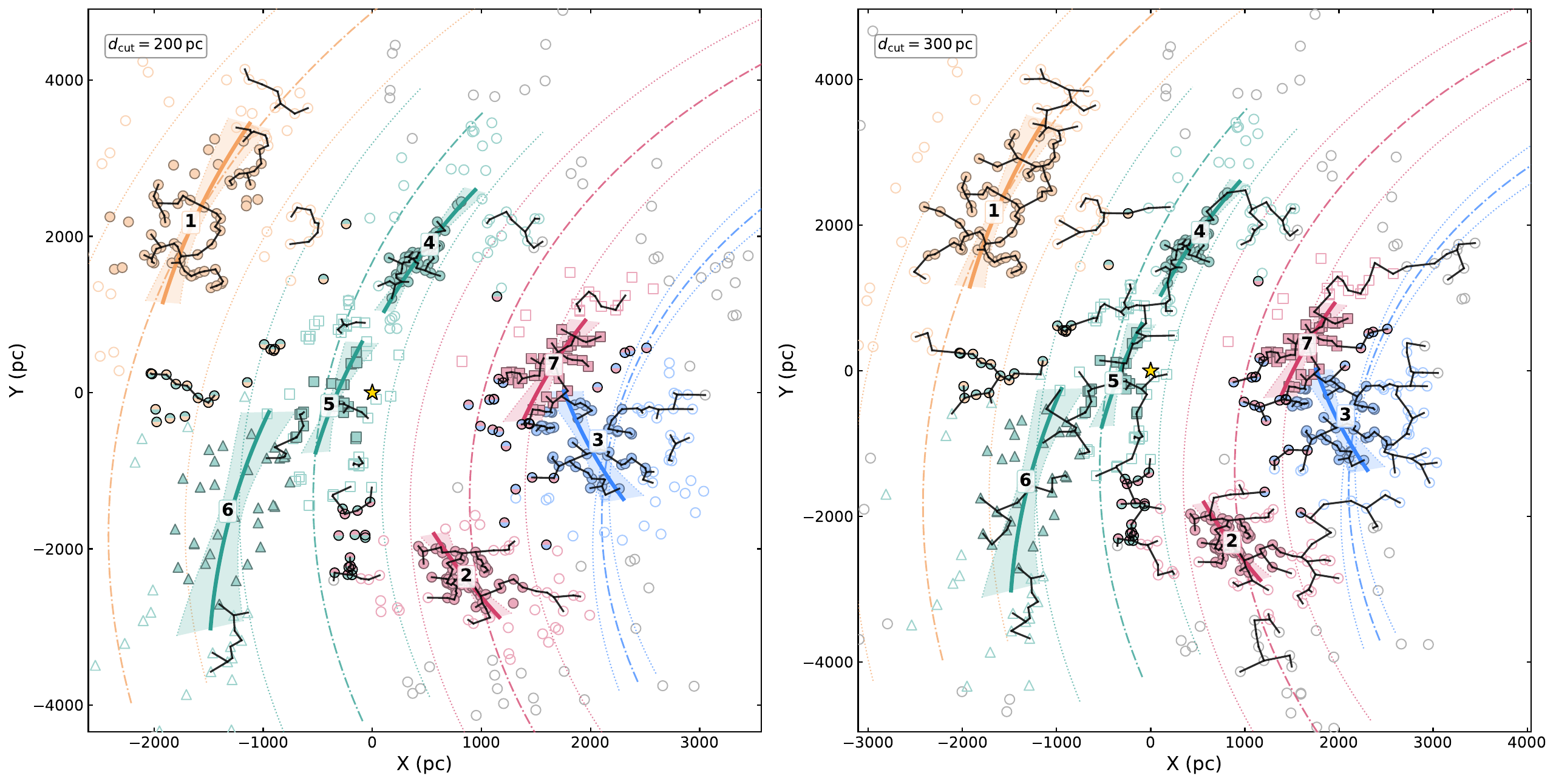}
    \includegraphics[width=0.95\textwidth]{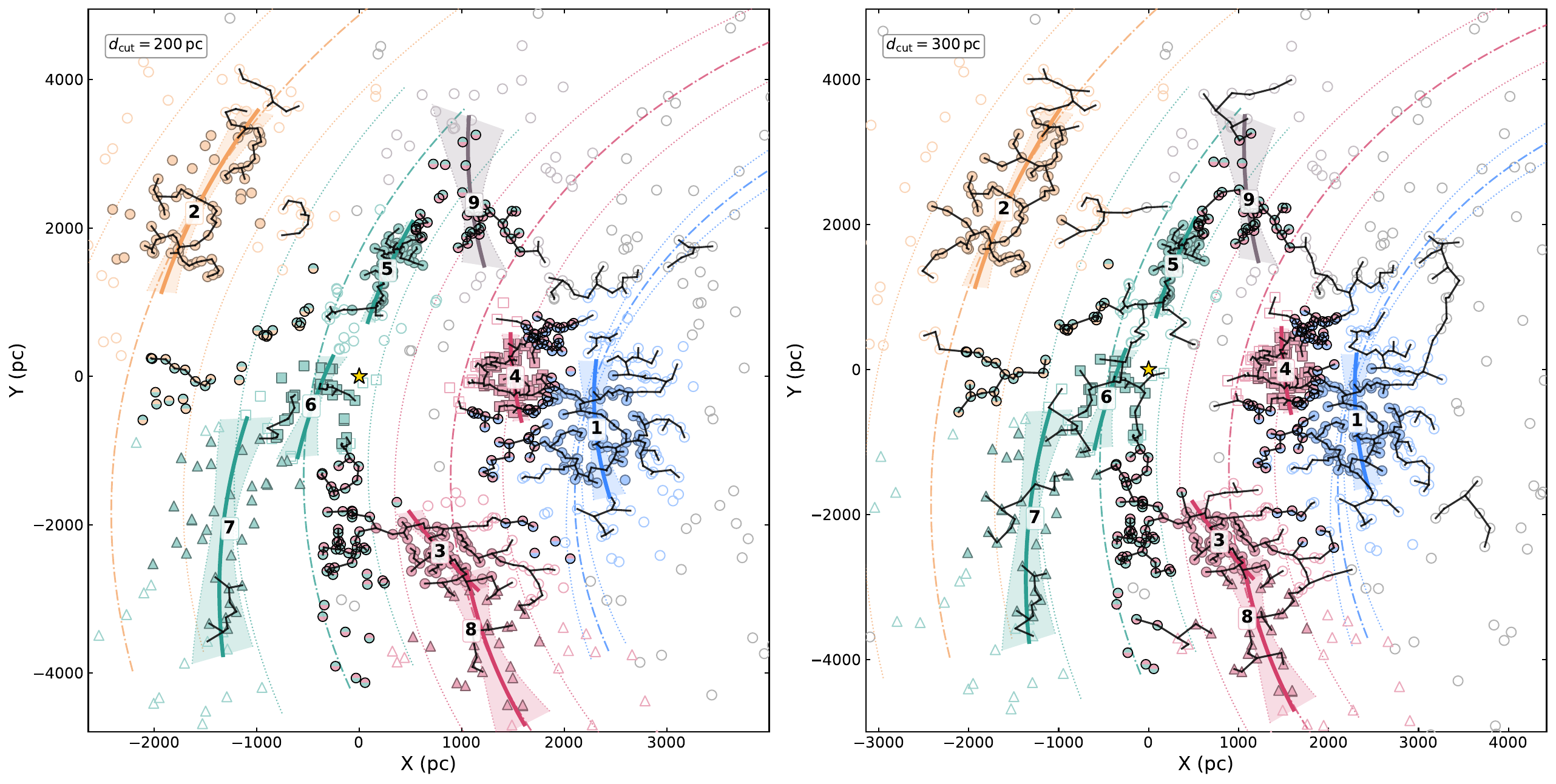}
    \caption{Minimum spanning tree connectivity in the heliocentric $(X,Y)$ plane. Top row: OC sample. Bottom row: OC+YSO sample. Left panels show $d_{\rm cut}=200\,{\rm pc}$ and right panels show $d_{\rm cut}=300\,{\rm pc}$. Only connected components with at least $N_{\rm min}=7$ nodes are retained. Increasing the linking scale merges neighbouring branches into larger connected structures while preserving a segmented morphology. Black lines trace the retained MST branches, while the published spiral-arm loci and the BGMM local segment fits are overplotted for direct visual comparison. The numbered labels identify the BGMM segments and the yellow star marks the Sun. The MST itself remains independent of both the reference arms and the BGMM assignments. Each panel is independently framed around the nodes belonging to the retained branches at the corresponding $d_{\rm cut}$, in order to display the scale-dependent connectivity more clearly.}
    \label{fig:mst_xy}
\end{figure*}

In the MST analysis (Fig.~\ref{fig:mst_xy}), we adopt $d_{\rm cut}=200\,{\rm pc}$ and $300\,{\rm pc}$ with $N_{\rm min}=7$. Full robustness tests are provided in Appendix~\ref{app:robustness}. The baseline configuration yields a total of seven segments for OC and nine segments for OC+YSO.

\section{Results: density segmentation and connectivity of the local young-tracer distribution}
\label{sec:results}

\begin{figure*}[t]
    \centering
    \includegraphics[width=0.48\textwidth]{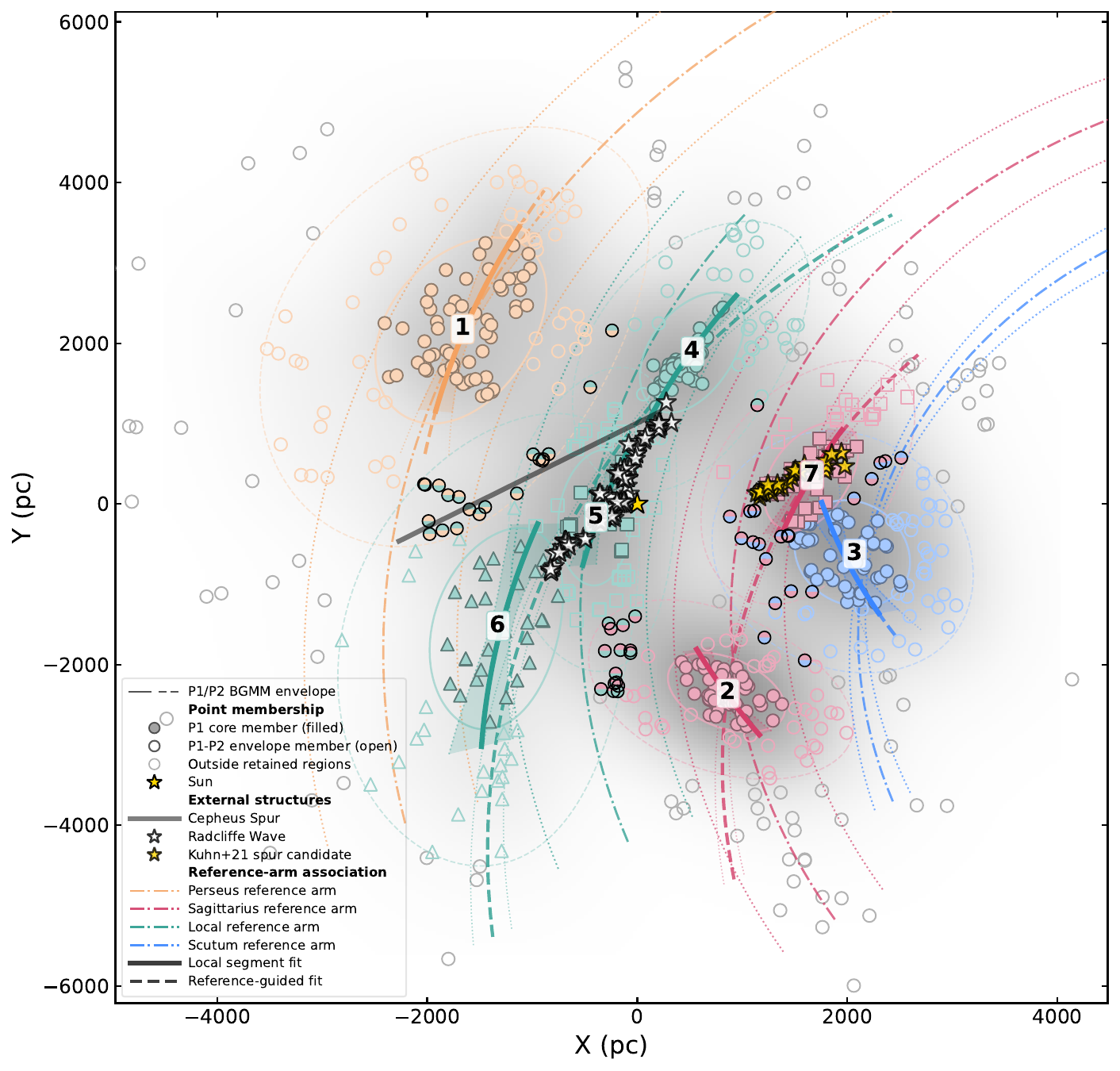}
    \includegraphics[width=0.48\textwidth]{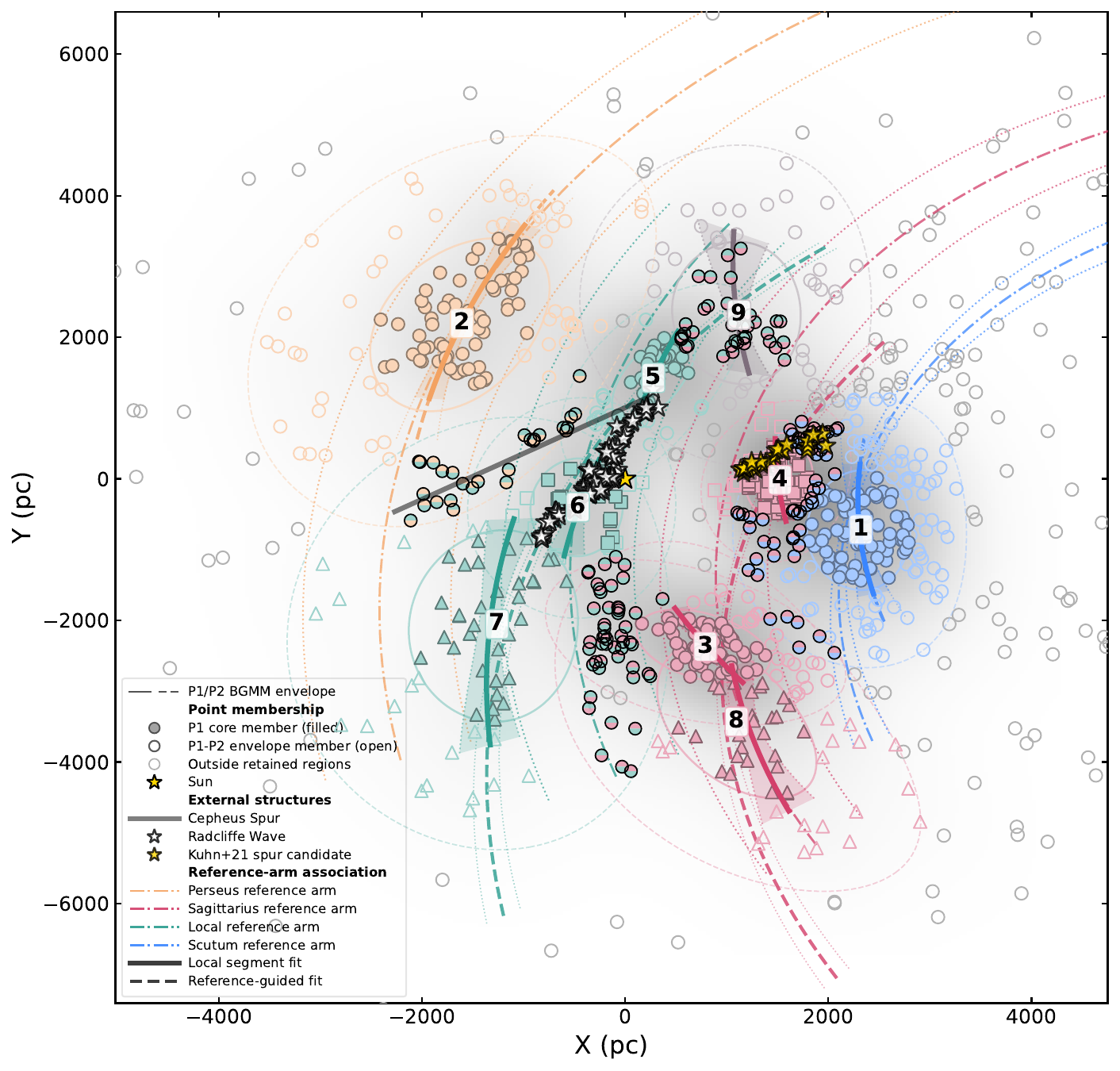}
    \caption{Reference-guided BGMM classification in the heliocentric $(X,Y)$ plane for the open-cluster-only sample (left panel) and the OC+YSO sample (right panel). The additional YSO-based groups increase the sampling density of recent star-forming structures and populate several local branches more clearly, while preserving the overall segmented morphology rather than producing continuous logarithmic ridges. Coloured symbols show the retained BGMM components, while grey points represent the full tracer distribution. The background indicates the smoothed density field. Published spiral arms are shown only as external reference curves and are not used to define the segmentation. Filled symbols denote $P_1$ core members, open coloured symbols denote $P_2\setminus P_1$ envelope members, and light-grey open circles denote tracers outside the retained classification regions. The panels are framed around the retained local structures; the full spatial extent of the input catalogues is shown separately in Fig.~\ref{fig:input_catalogues_xy}. The yellow star marks the position of the Sun. Thick coloured lines show the fitted local segment orientations, while dashed black lines show the corresponding reference-guided fits.}
    \label{fig:bgmm_xy_reference}
\end{figure*}

The results below describe the morphology obtained from the density and graph-based diagnostics. The reference spiral arms are used only as a posteriori geometrical references and do not enter the segmentation procedure.

\subsection{Reference-guided morphology in the OC and OC+YSO samples}

The OC sample provides the conservative open-cluster-only view of the young-tracer distribution, while the OC+YSO sample includes the Kuhn et al. YSO-based groups and provides a denser sampling of recent clustered star formation.

Figure~\ref{fig:bgmm_xy_reference} shows the retained BGMM components
in the heliocentric $(X,Y)$ plane for the OC and OC+YSO samples in the left and right panels, respectively. The coloured components correspond to density-supported segments identified before any reference-arm association is assigned. The logarithmic spiral arms from the literature are shown afterwards only as reference loci.

In both samples, several segments are located close to the Perseus, Local, Sagittarius, and Scutum reference-arm regions. However, the correspondence is not one-to-one: some components lie between neighbouring reference loci or overlap with more than one arm envelope. The Local--Sagittarius region provides the clearest example, where the BGMM identifies an intermediate segment that is not uniquely associated with either reference arm.

The symbol filling encodes BGMM membership rather than proximity to a published arm. Filled symbols correspond to the conservative $P_1$ cores, while open coloured symbols trace the outer $P_2 \setminus P_1$ envelopes. Light-grey open symbols are not retained as classified members, either because they fall outside the robust Stage~1 sample or because they lie outside the retained $P_2$ envelopes. Some of these unassigned tracers happen to lie close to a reference-arm curve in projection, but this proximity alone does not trigger membership because the reference arms are introduced only a posteriori and do not enter the BGMM classification.

\subsection{Local morphology relative to grand-design reference arms}

The logarithmic spiral arms provide a geometrical reference frame for the local segments. Table~\ref{tab:reference_guided_fits} compares the reference parameters of \citet{Castro-Ginard2021} with the local fits obtained from the BGMM components associated with each reference arm.

The comparison is performed locally: the fitted parameters summarise the geometry of the young-tracer segments associated with each arm region and are not intended as a new global optimisation of the Milky Way spiral pattern.

In Table~\ref{tab:reference_guided_fits}, $R^2_{\rm fit}$ denotes the coefficient of determination of the weighted linear fit in the $(\theta_G,\ln R_G)$ plane. The local fits are not equally constrained for all associations. The Local and Sagittarius segments show the strongest linear trends, with high $R^2_{\rm fit}$ values in both samples. Perseus is weakly constrained in the present local volume, while the Scutum-associated component does not define a significant linear relation in the $(\theta_G,\ln R_G)$ plane. The Scutum fit is therefore included only for reference.

The Local-arm-associated fit yields a larger pitch angle than the adopted logarithmic reference arm. This difference reflects the geometry of the young-tracer segment sampled in the local volume rather than a global revision of the Local Arm pitch angle. The fitted value for OC+YSO, $\psi=23.7^\circ\pm0.7^\circ$, characterises the local orientation of the young structure associated with the Local-arm region.

\subsection{MST connectivity and scale-dependent branch merging}

The MST provides an independent spatial view of the young-tracer distribution in the heliocentric $(X,Y)$ plane. It is constructed without using the BGMM segmentation or any spiral-arm assignment.

After pruning edges longer than $d_{\rm cut}$ and retaining only connected components with at least $N_{\rm min}=7$ nodes, the connectivity changes systematically with scale. For the OC sample, the number of retained branches decreases from 20 at $d_{\rm cut}=200\,{\rm pc}$ to 9 at $d_{\rm cut}=300\,{\rm pc}$, while the number of connected tracers increases from 416 to 596. For OC+YSO, the corresponding values are 18 and 9 branches, with 704 and 864 connected tracers.

At the smaller pruning scale, the MST separates the distribution into several local branches. At the larger scale, neighbouring branches merge into extended structures, but the resulting network remains fragmented rather than collapsing into a single continuous arm.

The main MST branches overlap spatially with the overdensities identified by the BGMM, despite the two methods being constructed in different coordinate spaces and using different criteria. A representative case is the Local--Sagittarius region, where the intermediate BGMM segment coincides with a region where neighbouring MST branches connect as the linking scale increases.

\subsection{Role of the different tracer samples}

The comparison between OC and OC+YSO provides a test of the stability of the detected morphology. The segmented distribution is already present in the OC-only sample, showing that the main result does not depend exclusively on the addition of YSO-based groups.

The OC+YSO sample provides a denser sampling of the same young clustered structures and improves the visibility of local branches and intermediate regions. We therefore use OC as the conservative case and OC+YSO as the main representation of the local young-tracer morphology.

\section{Discussion}
\label{sec:discussion}

The results indicate that the young spiral structure in the solar neighbourhood is better described as a set of local density-supported segments with partial connectivity than as a small number of smooth continuous logarithmic ridges. The classical Perseus, Local, Sagittarius, and Scutum arms remain useful as a large-scale geometrical framework, but they do not capture the full complexity of the young clustered population traced on few-hundred-parsec scales.

A central aspect of this analysis is that the reference arms are not used to define the morphology. The BGMM segmentation is performed independently in the $(\theta_G,\ln R_G)$ plane, and the reference-arm comparison is introduced only afterwards. This separation allows the data to reveal structures that are not necessarily aligned with a predefined grand-design pattern. Some segments are naturally associated with classical arm loci, whereas others appear as intermediate, locally tilted, or bridge-like structures.

The agreement between BGMM and MST provides an additional constraint on this interpretation. The two diagnostics operate in different spaces and use different information: the BGMM identifies density-supported components in spiral coordinates, while the MST measures spatial connectivity directly in the heliocentric plane. Their agreement therefore indicates that the detected morphology is not only a consequence of the chosen coordinate representation. In particular, the Local--Sagittarius region is highlighted by both approaches: the BGMM identifies an intermediate component between the reference-arm loci, and the MST connects neighbouring branches through the same region when the allowed linking scale is increased. This makes it one of the strongest morphological candidates for a bridge-like or inter-arm structure in the present sample.

The MST scale dependence further illustrates the hierarchical nature of the local pattern. At small linking lengths, the young tracers separate into several local branches, while increasing the pruning scale merges some neighbouring structures. However, even at the larger connectivity scale explored here, the distribution does not collapse into a small number of continuous arm-like features. The parameter $d_{\rm cut}$ should therefore be interpreted as a morphological connectivity scale rather than as a physical arm width or dynamical length scale.

The local spiral fits provide another indication that young structures do not necessarily follow the same geometry as large-scale reference arms. The Local-arm-associated segment is the clearest example: the fitted pitch angle for the OC+YSO sample, $\psi=23.7^\circ\pm0.7^\circ$, is larger than the adopted reference value, but similar to that obtained by \citet{Vazquez2008}. This difference should not be interpreted as a revision of the global Local Arm pitch angle. Instead, it describes the local orientation of the young-tracer structure sampled in the solar neighbourhood. Similar high-inclination local features have been reported for nearby young structures, including the Radcliffe Wave and Orion-arm related features \citep{Konietzka2024}, suggesting that local young populations may contain tilted substructures embedded within a more complex spiral environment.

The comparison between OC and OC+YSO is important in this context. The segmented morphology is already present in the open-cluster-only sample, indicating that the main result is not generated solely by the inclusion of YSO-based groups. The OC+YSO sample provides a denser representation of the same recent star-formation structures and improves the visibility of local branches and intermediate regions. The two samples therefore provide complementary views of the same morphological pattern.

Regarding interstellar extinction, line-of-sight dust extinction does not significantly distort the spatial distribution of young open clusters within $4.5~\text{kpc}$, as bright upper main-sequence stars remain well within the Gaia detection threshold even under high extinction ($A_V \approx 6~\text{mag}$). 
Within the central few-kiloparsec volume, the youngest and more highly populated open clusters are expected to be less affected by Gaia incompleteness than older or lower-mass systems. However, extinction, crowding, distance, cluster mass, and catalogue-specific selection remain relevant. We therefore do not claim that the observed distribution is complete. The persistence of the main segmented morphology in the OC and OC+YSO samples nevertheless indicates that it is not produced solely by the addition of the YSO groups, although selection effects may still affect the contrast and detailed extent of individual segments.

The interpretation presented here remains morphological. The detected segments, bridges, and connectivity patterns do not by themselves determine whether the observed structure originates from the formation mechanism of the spiral pattern, from transient spiral features, from differential shear, or from later perturbations such as feedback or local dynamical evolution. Addressing these possibilities requires additional information beyond the spatial distribution analysed here, particularly ages, velocities, and vertical structure.

While density-wave theory \citep{Lin1964} accounts for coherent grand-design arms ($m = 2$), models based on local instabilities, swing amplification, or stochastic star formation (SSPSF) naturally generate ragged, high-order ($m \gg 2$) flocculent patterns \citep[e.g.][]{Gerola1978, Athanassoula1987}. At the spatial scales and for the clustered tracers analysed here, the data are not well described by a purely smooth grand-design pattern and are instead compatible with a hybrid morphology in which major spiral arms coexist with local flocculent branches, spurs, and inter-arm bridges \citep[e.g.][]{Elmegreen1987, Dobbs2006}. However, this morphological compatibility does not identify the underlying dynamical mechanism, which may involve a combination of global and local processes.

Independent young-population and interstellar-medium tracers provide a useful qualitative comparison. Maser parallaxes of high-mass star-forming regions support a large-scale four-arm description of the Milky Way, but also require additional arm segments, spurs, and local departures from simple log-periodic loci \citep{Reid2019}; in the solar neighbourhood, maser astrometry has also revealed a substantial Local Arm and a spur branching between the Local and Sagittarius arms \citep{Xu2016}. Gaia-based maps of OB stars likewise show pronounced local concentrations and substructure in the distribution of massive young stars \citep{Pantaleoni2021}. Classical Cepheids, although typically older than the tracers used here, form spatial groups that are broadly consistent with spiral features identified from masers and OB(A)/upper-main-sequence stars \citep{Lemasle2022}. On the gaseous side, H{\sc I} surface-density maps reveal a non-axisymmetric multi-armed disc with strong local variations \citep{Levine2006}, while recent three-dimensional dust and gas reconstructions show elongated and fragmented nearby structures that need not coincide exactly with ideal logarithmic arms \citep{Kormann2026,Soler2025,Shen2025}. These comparisons do not imply a one-to-one correspondence, because the tracers span different ages and selection functions, but they support a picture in which large-scale spiral organisation coexists with local segments, spurs, bridges, and tracer-dependent substructure. A homogeneous cross-tracer analysis remains beyond the scope of the present paper. A direct, consistently selected comparison with O/OB stars, H\,{\sc ii} regions, and molecular-gas tracers is therefore left for future work.

Future Gaia data releases and infrared astrometric surveys will improve the phase-space characterisation of young clustered populations, especially in regions affected by extinction. Such information will be required to determine whether the locally segmented pattern identified here represents a transient arrangement of star formation or a long-lived structural component of the Galactic disc.

Within the limits of the present analysis, the main conclusion is therefore morphological: the young clustered population in the solar neighbourhood traces a segmented, partially connected, and locally irregular spiral pattern rather than a set of smooth grand-design arms.

\section{Conclusions}
\label{sec:conclusions}

We have analysed the local spiral morphology traced by young open clusters and YSO-based star-forming groups in the solar neighbourhood. Rather than imposing a global spiral-arm model, we have used a density-supported segmentation in the $(\theta_G,\ln R_G)$ plane and an independent connectivity analysis in the heliocentric $(X,Y)$ plane to characterise how young clustered tracers organise themselves.

The main conclusions of this work are:

\begin{enumerate}

\item The young clustered population in the solar neighbourhood is better described as a set of local density-supported segments than as a small number of smooth continuous logarithmic ridges. The BGMM decomposition identifies several spatially coherent structures with limited angular extent rather than a single continuous arm pattern.

\item Classical logarithmic spiral arms remain useful as a large-scale reference frame, but they do not fully describe the local morphology traced by young populations. Several segments are compatible with the Perseus, Local, Sagittarius, and Scutum loci, while others occupy intermediate regions and are more naturally interpreted as candidate bridges, spurs, bifurcations, or inter-arm concentrations.

\item The segmented morphology is already present in the OC-only sample, showing that it is not produced solely by the inclusion of YSO-based groups. The OC+YSO sample provides a denser sampling of recent clustered star formation and improves the visibility of local connections, while preserving the same overall morphological picture.

\item The MST analysis independently supports the segmented character of the distribution. Changing the pruning scale from $d_{\rm cut}=200\,{\rm pc}$ to $300\,{\rm pc}$ merges neighbouring branches but does not produce a small number of continuous arm-like structures. The connectivity therefore reveals a hierarchy of local structures rather than a single coherent ridge.

\item The Local--Sagittarius region is the clearest example of a possible connection between local segments. A BGMM component appears between the reference-arm loci, and MST branches connect through the same region when larger spatial linking scales are allowed. This agreement between independent diagnostics makes the region a strong morphological candidate for a bridge-like or inter-arm structure, although the present analysis does not establish dynamical coherence.

\item The local spiral fits show that young structures can depart significantly from the geometry of large-scale reference arms. In particular, the Local-arm-associated segment has a larger pitch angle than the adopted reference model. This should be interpreted as the orientation of a local young structure rather than as a revision of the global Local Arm geometry.

\item  The local morphology is compatible with a hybrid picture in which large-scale spiral structures coexist with shorter, locally flocculent segments. The present analysis does not by itself determine the global spiral class or dynamical origin of the Milky Way pattern.

\end{enumerate}

The result established here is therefore primarily morphological: nearby young tracers form a segmented, partially connected, and locally irregular spiral pattern rather than a set of smooth grand-design arms. The present analysis does not determine whether this structure originates from the formation mechanism of spiral patterns, from transient spiral features, or from subsequent evolution driven by shear, feedback, and local dynamics.

A natural next step is to combine this morphological framework with ages, velocities, and vertical phase-space information to test whether individual segments and candidate bridges represent physically coherent populations. Catalogue-dependent stability tests and improved cross-identification will also be required to quantify the sensitivity of individual components to the adopted tracer catalogues.

Future astrometric and spectroscopic surveys may allow this picture to be extended from morphology to dynamics. An interesting possibility is that the segmented pattern revealed here is not merely a disrupted version of a classical spiral arm, but instead reflects a hierarchical mode of star formation in which local overdensities, bridges, and short spiral-like segments emerge as coupled elements of a larger Galactic disc response. Testing this hypothesis will require combining spatial morphology with the kinematics and evolutionary state of the young populations.

\section*{Data availability}

The open-cluster catalogues used in this work are publicly available from the original publications and associated data services. The YSO-based star-forming groups are taken from the catalogue of \citet{Kuhn2021}. The derived working tracer catalogue, BGMM classifications, and MST branch memberships are provided in electronic Tables~3--5, which have been submitted to the CDS.

\begin{acknowledgements}
We thank the referee for their careful and constructive comments, which have helped to improve both the content and presentation of this paper. This work has been partially funded by the Spanish MCIN/AEI grants PID2022-136640NB-C21, and PID2025-175328NB-I00. The authors acknowledge financial support from the Severo Ochoa grant CEX2021-001131-S, funded by MCIN/AEI/10.13039/501100011033. This research has used TOPCAT \citep{TopCat2005}.

This work presents results based on data from the European Space Agency (ESA) mission Gaia. The Gaia data are processed by the Gaia Data Processing and Analysis Consortium (DPAC). Funding for the DPAC is provided by national institutions, particularly those participating in the Gaia Multilateral Agreement. The Gaia mission website is \url{https://www.cosmos.esa.int/gaia}, and the Gaia archive website is \url{https://archives.esac.esa.int/gaia}.
\end{acknowledgements}

\bibliographystyle{bibtex/aa}
\bibliography{REFERENCES.bib}

@ARTICLE{Morgan1952,
  author  = {{Morgan}, W. W. and {Sharpless}, S. and {Osterbrock}, D.},
  title   = {Some features of Galactic structure in the neighborhood of the Sun},
  journal = {AJ},
  year    = {1952},
  volume  = {57},
  pages   = {3}
}

@ARTICLE{Morgan1953,
  author  = {{Morgan}, W. W. and {Whitford}, A. E. and {Code}, A. D.},
  title   = {Studies in Galactic structure. I. A preliminary determination of the space distribution of the blue giants},
  journal = {ApJ},
  year    = {1953},
  volume  = {118},
  pages   = {318}
}

@ARTICLE{Lin1964,
  author  = {{Lin}, C. C. and {Shu}, F. H.},
  title   = {On the spiral structure of disk galaxies},
  journal = {ApJ},
  year    = {1964},
  volume  = {140},
  pages   = {646}
}

@ARTICLE{Lin1966,
  author  = {{Lin}, C. C. and {Shu}, F. H.},
  title   = {On the spiral structure of disk galaxies, II. Outline of a theory of density waves},
  journal = {Proceedings of the National Academy of Science},
  year    = {1966},
  volume  = {55},
  pages   = {229}
}

@ARTICLE{Levine2006,
  author  = {{Levine}, E. S. and {Blitz}, L. and {Heiles}, C.},
  title   = {The spiral structure of the outer Milky Way in hydrogen},
  journal = {ApJ},
  year    = {2006},
  volume  = {643},
  pages   = {881--896}
}

@ARTICLE{Xu2013,
  author  = {{Xu}, Y. and {Li}, J. J. and {Reid}, M. J. and others},
  title   = {On the nature of the Local Arm of the Milky Way},
  journal = {ApJ},
  year    = {2013},
  volume  = {769},
  pages   = {15}
}

@ARTICLE{Pettitt2015,
  author  = {{Pettitt}, A. R. and {Dobbs}, C. L. and {Acreman}, D. M. and {Bate}, M. R.},
  title   = {The morphology of spiral arms in simulated Milky Way-like galaxies},
  journal = {MNRAS},
  year    = {2015},
  volume  = {449},
  pages   = {3911--3927}
}

@ARTICLE{Xu2016,
  author  = {{Xu}, Y. and {Reid}, M. J. and {Dame}, T. M. and others},
  title   = {The local spiral structure of the Milky Way},
  journal = {Science Advances},
  year    = {2016},
  volume  = {2},
  pages   = {e1600878}
}

@ARTICLE{Dias2021,
  author  = {{Dias}, W. S. and {Monteiro}, H. and {Moitinho}, A. and {L\'epine}, J. R. D. and {Carraro}, G. and {Paunzen}, E. and {Alessi}, B. and {Villela}, L.},
  title   = {Updated parameters of 1743 open clusters based on Gaia DR2},
  journal = {MNRAS},
  year    = {2021},
  volume  = {504},
  pages   = {356--371},
  doi     = {10.1093/mnras/stab770}
}

@ARTICLE{Kuhn2021,
  author  = {{Kuhn}, M. A. and {Benjamin}, R. A. and {Zucker}, C. and others},
  title   = {A high pitch angle structure in the Sagittarius Arm},
  journal = {A\&A},
  year    = {2021},
  volume  = {651},
  pages   = {L10}
}

@ARTICLE{Castro-Ginard2021,
  author  = {{Castro-Ginard}, A. and {McMillan}, P. J. and {Luri}, X. and others},
  title   = {Mapping the spiral structure of the Milky Way with open clusters},
  journal = {A\&A},
  year    = {2021},
  volume  = {652},
  pages   = {A162}
}

@ARTICLE{Hao2022,
  author  = {{Hao}, C. J. and {Xu}, Y. and {Wu}, Z. Y. and {Lin}, Z. H. and {Liu}, D. J. and {Li}, Y. J.},
  title   = {Newly detected open clusters in the Galactic disk using Gaia EDR3},
  journal = {A\&A},
  year    = {2022},
  volume  = {660},
  pages   = {A4},
  doi     = {10.1051/0004-6361/202243091}
}

@ARTICLE{Martinez2022,
  author  = {{Mart\'inez-Medina}, L. and {P\'erez-Villegas}, A. and {Peimbert}, A.},
  title   = {Vertical perturbations and spiral structure in the Galactic disc},
  journal = {MNRAS},
  year    = {2022},
  volume  = {512},
  pages   = {1574--1587}
}

@ARTICLE{Soler2025,
  author  = {{Soler}, J. D. and {Molinari}, S. and {Glover}, S. C. O. and others},
  title   = {The structure of the local interstellar medium},
  journal = {A\&A},
  year    = {2025},
  volume  = {695},
  pages   = {A222}
}

@ARTICLE{Shen2025,
  author  = {{Shen}, X. J. and {Hou}, L. G. and {Liu}, H. L. and {Gao}, X. Y.},
  title   = {The Local Arm as a large-pitch-angle segment traced by young objects},
  journal = {A\&A},
  year    = {2025},
  volume  = {696},
  pages   = {A67}
}

@ARTICLE{Zari2025,
  author  = {{Zari}, E. and {Villase\~nor}, J. and {Kounkel}, M. and others},
  title   = {Young stars and phase-space substructure in the solar neighbourhood},
  journal = {A\&A},
  year    = {2025},
  volume  = {703},
  pages   = {A303}
}

@ARTICLE{Liu2025,
  author  = {{Liu}, X. and {He}, Z. and {Luo}, Y. and {Wang}, K.},
  title   = {Age gradients and young open clusters in the Galactic disc},
  journal = {MNRAS},
  year    = {2025},
  volume  = {537},
  pages   = {2403--2418}
}

@ARTICLE{Viscasilla2025,
  author  = {{Viscasillas V\'azquez}, C. and {Magrini}, L. and {Spitoni}, E. and others},
  title   = {Young tracers and Galactic disc structure},
  journal = {arXiv e-prints},
  year    = {2025},
  pages   = {arXiv:2504.10000},
  eprint  = {2504.10000},
  archivePrefix = {arXiv},
  primaryClass  = {astro-ph.GA}
}

@INPROCEEDINGS{TopCat2005,
  author    = {{Taylor}, M. B.},
  title     = {{TOPCAT} \\& {STIL}: Starlink table/VOTable processing software},
  booktitle = {Astronomical Data Analysis Software and Systems XIV},
  editor    = {{Shopbell}, P. and {Britton}, M. and {Ebert}, R.},
  series    = {ASP Conference Series},
  volume    = {347},
  year      = {2005},
  pages     = {29}
}

@ARTICLE{Reid2019,
       author = {{Reid}, M.~J. and {Menten}, K.~M. and {Brunthaler}, A. and {Zheng}, X.~W. and {Dame}, T.~M. and {Xu}, Y. and {Li}, J. and {Sakai}, N. and {Wu}, Y. and {Immer}, K. and {Zhang}, B. and {Sanna}, A. and {Moscadelli}, L. and {Rygl}, K.~L.~J. and {Bartkiewicz}, A. and {Hu}, B. and {Quiroga-Nu{\~n}ez}, L.~H. and {van Langevelde}, H.~J.},
        title = "{Trigonometric Parallaxes of High-mass Star-forming Regions: Our View of the Milky Way}",
      journal = {\apj},
         year = 2019,
        month = nov,
       volume = {885},
       number = {2},
          eid = {131},
        pages = {131},
          doi = {10.3847/1538-4357/ab4a11},
archivePrefix = {arXiv},
       eprint = {1910.03357},
 primaryClass = {astro-ph.GA},
       adsurl = {https://ui.adsabs.harvard.edu/abs/2019ApJ...885..131R}
}

@ARTICLE{Roberts1970,
       author = {{Roberts}, Jr., William W. and {Yuan}, C.},
        title = "{Application of the Density-Wave Theory to the Spiral Structure of the Milky way System. III. Magnetic Field: Large-Scale Hydromagnetic Shock Formation}",
      journal = {\apj},
         year = 1970,
        month = sep,
       volume = {161},
        pages = {887},
          doi = {10.1086/150592},
       adsurl = {https://ui.adsabs.harvard.edu/abs/1970ApJ...161..887R}
}

@ARTICLE{Roberts1969,
       author = {{Roberts}, W.~W.},
        title = "{Large-Scale Shock Formation in Spiral Galaxies and its Implications on Star Formation}",
      journal = {\apj},
         year = 1969,
        month = oct,
       volume = {158},
        pages = {123},
          doi = {10.1086/150177},
       adsurl = {https://ui.adsabs.harvard.edu/abs/1969ApJ...158..123R}
}

@ARTICLE{Davies1960,
       author = {{Davies}, R.~D.},
        title = "{A study of neutral hydrogen in the solar neighbourhood of the Milky Way}",
      journal = {\mnras},
         year = 1960,
        month = jan,
       volume = {120},
        pages = {483},
          doi = {10.1093/mnras/120.5.483},
       adsurl = {https://ui.adsabs.harvard.edu/abs/1960MNRAS.120..483D}
}

@ARTICLE{Becker1956,
       author = {{Becker}, Wilhelm},
        title = "{An analysis of the Milky Way}",
      journal = {Vistas in Astronomy},
         year = 1956,
        month = jan,
       volume = {2},
       number = {1},
        pages = {1515-1522},
          doi = {10.1016/0083-6656(56)90081-2},
       adsurl = {https://ui.adsabs.harvard.edu/abs/1956VA......2.1515B}
}

@ARTICLE{Georgelin1976,
       author = {{Georgelin}, Y.~M. and {Georgelin}, Y.~P.},
        title = "{The spiral structure of our Galaxy determined from H II regions.}",
      journal = {\aap},
         year = 1976,
        month = may,
       volume = {49},
        pages = {57-79},
       adsurl = {https://ui.adsabs.harvard.edu/abs/1976A&A....49...57G}
}

@ARTICLE{Kormann2026,
       author = {{Kormann}, Lilly A. and {Alves}, Jo{\~a}o and {Pantaleoni Gonz{\'a}lez}, Michelangelo and {Swiggum}, Cameren and {En{\ss}lin}, Torsten A. and {Edenhofer}, Gordian},
        title = "{The superclouds of the local Milky Way}",
      journal = {\aap},
         year = 2026,
        month = feb,
       volume = {706},
          eid = {A161},
        pages = {A161},
          doi = {10.1051/0004-6361/202556469},
archivePrefix = {arXiv},
       eprint = {2507.14883},
 primaryClass = {astro-ph.GA},
       adsurl = {https://ui.adsabs.harvard.edu/abs/2026A&A...706A.161K}
}

@ARTICLE{Alfaro2025,
       author = {{Alfaro}, Emilio J. and {S{\'a}nchez-Gil}, M. Carmen and {Elmegreen}, Bruce},
        title = "{Vertical kinematics of the young Galactic clusters}",
      journal = {\mnras},
         year = 2025,
        month = mar,
       volume = {537},
       number = {4},
        pages = {3066-3077},
          doi = {10.1093/mnras/staf155},
archivePrefix = {arXiv},
       eprint = {2501.05187},
 primaryClass = {astro-ph.GA},
       adsurl = {https://ui.adsabs.harvard.edu/abs/2025MNRAS.537.3066A}
}

@ARTICLE{Alfaro2022,
       author = {{Alfaro}, Emilio J. and {Jim{\'e}nez}, Manuel and {S{\'a}nchez-Gil}, M. Carmen and {S{\'a}nchez}, N{\'e}stor and {Gonz{\'a}lez}, Marta and {Ma{\'\i}z Apell{\'a}niz}, Jes{\'u}s},
        title = "{Topography of the Young Galactic Disk: Spatial and Kinematic Patterns of Clustered Star Formation in the Solar Neighborhood}",
      journal = {\apj},
         year = 2022,
        month = oct,
       volume = {937},
       number = {2},
          eid = {114},
        pages = {114},
          doi = {10.3847/1538-4357/ac8b0c},
archivePrefix = {arXiv},
       eprint = {2206.11313},
 primaryClass = {astro-ph.GA},
       adsurl = {https://ui.adsabs.harvard.edu/abs/2022ApJ...937..114A}
}

@ARTICLE{Scoville1975,
       author = {{Scoville}, N.~Z. and {Solomon}, P.~M.},
        title = "{Molecular clouds in the Galaxy.}",
      journal = {\apjl},
         year = 1975,
        month = jul,
       volume = {199},
        pages = {L105-L109},
          doi = {10.1086/181859},
       adsurl = {https://ui.adsabs.harvard.edu/abs/1975ApJ...199L.105S}
}

@ARTICLE{Cohen1980,
       author = {{Cohen}, R.~S. and {Cong}, H. and {Dame}, T.~M. and {Thaddeus}, P.},
        title = "{Molecular clouds and galactic spiral structure}",
      journal = {\apjl},
         year = 1980,
        month = jul,
       volume = {239},
        pages = {L53-L56},
          doi = {10.1086/183290},
       adsurl = {https://ui.adsabs.harvard.edu/abs/1980ApJ...239L..53C}
}

@ARTICLE{ Crampton1975,
       author = {{Crampton}, D. and {Georgelin}, Y.~M.},
        title = "{The distribution of optical H II regions in our Galaxy.}",
      journal = {\aap},
         year = 1975,
        month = may,
       volume = {40},
        pages = {317-321},
       adsurl = {https://ui.adsabs.harvard.edu/abs/1975A&A....40..317C}
}

@ARTICLE{VdHulst1954,
       author = {{van de Hulst}, H.~C. and {Muller}, C.~A. and {Oort}, J.~H.},
        title = "{The spiral structure of the outer part of the Galactic System derived from the hydrogen emission at 21 cm wavelength}",
      journal = {\bain},
         year = 1954,
        month = may,
       volume = {12},
        pages = {117},
       adsurl = {https://ui.adsabs.harvard.edu/abs/1954BAN....12..117V}
}

@ARTICLE{Vallee2017,
       author = {{Vall{\'e}e}, Jacques P.},
        title = "{A guided map to the spiral arms in the galactic disk of the Milky Way}",
      journal = {The Astronomical Review},
         year = 2017,
        month = oct,
       volume = {13},
       number = {3-4},
        pages = {113-146},
          doi = {10.1080/21672857.2017.1379459},
archivePrefix = {arXiv},
       eprint = {1711.05228},
 primaryClass = {astro-ph.GA},
       adsurl = {https://ui.adsabs.harvard.edu/abs/2017AstRv..13..113V}
}

@ARTICLE{Vallee2014,
       author = {{Vall{\'e}e}, Jacques P.},
        title = "{The Spiral Arms of the Milky Way: The Relative Location of Each Different Arm Tracer within a Typical Spiral Arm Width}",
      journal = {\aj},
         year = 2014,
        month = jul,
       volume = {148},
       number = {1},
          eid = {5},
        pages = {5},
          doi = {10.1088/0004-6256/148/1/5},
       adsurl = {https://ui.adsabs.harvard.edu/abs/2014AJ....148....5V}
}

@ARTICLE{Xu2018,
       author = {{Xu}, Y. and {Bian}, S.~B. and {Reid}, M.~J. and {Li}, J.~J. and {Zhang}, B. and {Yan}, Q.~Z. and {Dame}, T.~M. and {Menten}, K.~M. and {He}, Z.~H. and {Liao}, S.~L. and {Tang}, Z.~H.},
        title = "{A comparison of the local spiral structure from Gaia DR2 and VLBI maser parallaxes}",
      journal = {\aap},
         year = 2018,
        month = aug,
       volume = {616},
          eid = {L15},
        pages = {L15},
          doi = {10.1051/0004-6361/201833407},
archivePrefix = {arXiv},
       eprint = {1807.00315},
 primaryClass = {astro-ph.GA},
       adsurl = {https://ui.adsabs.harvard.edu/abs/2018A&A...616L..15X}
}

@ARTICLE{Lemasle2022,
       author = {{Lemasle}, B. and {Lala}, H.~N. and {Kovtyukh}, V. and {Hanke}, M. and {Prudil}, Z. and {Bono}, G. and {Braga}, V.~F. and {da Silva}, R. and {Fabrizio}, M. and {Fiorentino}, G. and {Fran{\c{c}}ois}, P. and {Grebel}, E.~K. and {Kniazev}, A.},
        title = "{Tracing the Milky Way warp and spiral arms with classical Cepheids}",
      journal = {\aap},
         year = 2022,
        month = dec,
       volume = {668},
          eid = {A40},
        pages = {A40},
          doi = {10.1051/0004-6361/202243273},
archivePrefix = {arXiv},
       eprint = {2209.02731},
 primaryClass = {astro-ph.GA},
       adsurl = {https://ui.adsabs.harvard.edu/abs/2022A&A...668A..40L}
}

@ARTICLE{Pantaleoni2021,
       author = {{Pantaleoni Gonz{\'a}lez}, M. and {Ma{\'\i}z Apell{\'a}niz}, J. and {Barb{\'a}}, R.~H. and {Reed}, B. Cameron},
        title = "{The Alma catalogue of OB stars - II. A cross-match with Gaia DR2 and an updated map of the solar neighbourhood}",
      journal = {\mnras},
         year = 2021,
        month = jun,
       volume = {504},
       number = {2},
        pages = {2968-2982},
          doi = {10.1093/mnras/stab688},
archivePrefix = {arXiv},
       eprint = {2103.02748},
 primaryClass = {astro-ph.SR},
       adsurl = {https://ui.adsabs.harvard.edu/abs/2021MNRAS.504.2968P}
}

@ARTICLE{Vazquez2008,
       author = {{V{\'a}zquez}, Ruben A. and {May}, Jorge and {Carraro}, Giovanni and {Bronfman}, Leonardo and {Moitinho}, Andr{\'e} and {Baume}, Gustavo},
        title = "{Spiral Structure in the Outer Galactic Disk. I. The Third Galactic Quadrant}",
      journal = {\apj},
         year = 2008,
        month = jan,
       volume = {672},
       number = {2},
        pages = {930-939},
          doi = {10.1086/524003},
archivePrefix = {arXiv},
       eprint = {0709.3973},
 primaryClass = {astro-ph},
       adsurl = {https://ui.adsabs.harvard.edu/abs/2008ApJ...672..930V}
}

@ARTICLE{Konietzka2024,
       author = {{Konietzka}, Ralf and {Goodman}, Alyssa A. and {Zucker}, Catherine and {Burkert}, Andreas and {Alves}, Jo{\~a}o and {Foley}, Michael and {Swiggum}, Cameren and {Koller}, Maria and {Miret-Roig}, N{\'u}ria},
        title = "{The Radcliffe Wave is oscillating}",
      journal = {\nat},
         year = 2024,
        month = apr,
       volume = {628},
       number = {8006},
        pages = {62-65},
          doi = {10.1038/s41586-024-07127-3},
archivePrefix = {arXiv},
       eprint = {2402.12596},
 primaryClass = {astro-ph.GA},
       adsurl = {https://ui.adsabs.harvard.edu/abs/2024Natur.628...62K}
}

@ARTICLE{Hunt2025,
       author = {{Hunt}, Emily L. and {Cantat-Gaudin}, Tristan and {Anders}, Friedrich and {Spina}, Lorenzo and {Cavallo}, Lorenzo and {Castro-Ginard}, Alfred and {Belokurov}, Vasily and {Brown}, Anthony G.~A. and {Casey}, Andrew R. and {Drimmel}, Ronald and {Fouesneau}, Morgan and {Reffert}, Sabine},
        title = "{The completeness of the open cluster census towards the Galactic anticentre}",
      journal = {\aap},
         year = 2025,
        month = jul,
       volume = {699},
          eid = {A273},
        pages = {A273},
          doi = {10.1051/0004-6361/202452614},
archivePrefix = {arXiv},
       eprint = {2506.18708},
 primaryClass = {astro-ph.GA},
       adsurl = {https://ui.adsabs.harvard.edu/abs/2025A&A...699A.273H}
}

@ARTICLE{Hunt2024,
       author = {{Hunt}, Emily L. and {Reffert}, Sabine},
        title = "{Improving the open cluster census. III. Using cluster masses, radii, and dynamics to create a cleaned open cluster catalogue}",
      journal = {\aap},
         year = 2024,
        month = jun,
       volume = {686},
          eid = {A42},
        pages = {A42},
          doi = {10.1051/0004-6361/202348662},
archivePrefix = {arXiv},
       eprint = {2403.05143},
 primaryClass = {astro-ph.GA},
       adsurl = {https://ui.adsabs.harvard.edu/abs/2024A&A...686A..42H}
}

@ARTICLE{Cantat2020,
       author = {{Cantat-Gaudin}, T. and {Anders}, F. and {Castro-Ginard}, A. and {Jordi}, C. and {Romero-G{\'o}mez}, M. and {Soubiran}, C. and {Casamiquela}, L. and {Tarricq}, Y. and {Moitinho}, A. and {Vallenari}, A. and {Bragaglia}, A. and {Krone-Martins}, A. and {Kounkel}, M.},
        title = "{Painting a portrait of the Galactic disc with its stellar clusters}",
      journal = {\aap},
         year = 2020,
        month = aug,
       volume = {640},
          eid = {A1},
        pages = {A1},
          doi = {10.1051/0004-6361/202038192},
archivePrefix = {arXiv},
       eprint = {2004.07274},
 primaryClass = {astro-ph.GA},
       adsurl = {https://ui.adsabs.harvard.edu/abs/2020A&A...640A...1C}
}

@ARTICLE{Cavallo2024,
       author = {{Cavallo}, Lorenzo and {Spina}, Lorenzo and {Carraro}, Giovanni and {Magrini}, Laura and {Poggio}, Eloisa and {Cantat-Gaudin}, Tristan and {Pasquato}, Mario and {Lucatello}, Sara and {Ortolani}, Sergio and {Schiappacasse-Ulloa}, Jose},
        title = "{Parameter Estimation for Open Clusters using an Artificial Neural Network with a QuadTree-based Feature Extractor}",
      journal = {\aj},
         year = 2024,
        month = jan,
       volume = {167},
       number = {1},
          eid = {12},
        pages = {12},
          doi = {10.3847/1538-3881/ad07e5},
archivePrefix = {arXiv},
       eprint = {2311.03009},
 primaryClass = {astro-ph.GA},
       adsurl = {https://ui.adsabs.harvard.edu/abs/2024AJ....167...12C}
}

@ARTICLE{CastroGinard2020,
       author = {{Castro-Ginard}, A. and {Jordi}, C. and {Luri}, X. and {{\'A}lvarez Cid-Fuentes}, J. and {Casamiquela}, L. and {Anders}, F. and {Cantat-Gaudin}, T. and {Mongui{\'o}}, M. and {Balaguer-N{\'u}{\~n}ez}, L. and {Sol{\`a}}, S. and {Badia}, R.~M.},
        title = "{Hunting for open clusters in Gaia DR2: 582 new open clusters in the Galactic disc}",
      journal = {\aap},
         year = 2020,
        month = mar,
       volume = {635},
          eid = {A45},
        pages = {A45},
          doi = {10.1051/0004-6361/201937386},
archivePrefix = {arXiv},
       eprint = {2001.07122},
 primaryClass = {astro-ph.GA},
       adsurl = {https://ui.adsabs.harvard.edu/abs/2020A&A...635A..45C}
}

@ARTICLE{Hunt2023,
       author = {{Hunt}, Emily L. and {Reffert}, Sabine},
        title = "{Improving the open cluster census. II. An all-sky cluster catalogue with Gaia DR3}",
      journal = {\aap},
         year = 2023,
        month = may,
       volume = {673},
          eid = {A114},
        pages = {A114},
          doi = {10.1051/0004-6361/202346285},
archivePrefix = {arXiv},
       eprint = {2303.13424},
 primaryClass = {astro-ph.GA},
       adsurl = {https://ui.adsabs.harvard.edu/abs/2023A&A...673A.114H}
}

@ARTICLE{Athanassoula1987,
       author = {{Athanassoula}, E. and {Bosma}, A. and {Papaioannou}, S.},
        title = "{Halo parameters of spiral galaxies.}",
      journal = {\aap},
         year = 1987,
        month = jun,
       volume = {179},
        pages = {23-40},
       adsurl = {https://ui.adsabs.harvard.edu/abs/1987A&A...179...23A}
}

@ARTICLE{Gerola1978,
       author = {{Gerola}, H. and {Seiden}, P.~E.},
        title = "{Stochastic star formation and spiral structure of galaxies.}",
      journal = {\apj},
         year = 1978,
        month = jul,
       volume = {223},
        pages = {129-139},
          doi = {10.1086/156243},
       adsurl = {https://ui.adsabs.harvard.edu/abs/1978ApJ...223..129G}
}

@ARTICLE{Elmegreen1987,
       author = {{Elmegreen}, Debra Meloy and {Elmegreen}, Bruce G.},
        title = "{Arm Classifications for Spiral Galaxies}",
      journal = {\apj},
         year = 1987,
        month = mar,
       volume = {314},
        pages = {3},
          doi = {10.1086/165034},
       adsurl = {https://ui.adsabs.harvard.edu/abs/1987ApJ...314....3E}
}

@ARTICLE{Dobbs2006,
       author = {{Dobbs}, C.~L. and {Bonnell}, I.~A.},
        title = "{Spurs and feathering in spiral galaxies}",
      journal = {\mnras},
         year = 2006,
        month = apr,
       volume = {367},
       number = {3},
        pages = {873-878},
          doi = {10.1111/j.1365-2966.2006.10146.x},
archivePrefix = {arXiv},
       eprint = {astro-ph/0602100},
 primaryClass = {astro-ph},
       adsurl = {https://ui.adsabs.harvard.edu/abs/2006MNRAS.367..873D}
}

@ARTICLE{Dobbs2018,
       author = {{Dobbs}, C.~L. and {Pettitt}, A.~R. and {Corbelli}, E. and {Pringle}, J.~E.},
        title = "{Simulations of the flocculent spiral M33: what drives the spiral structure?}",
      journal = {\mnras},
         year = 2018,
        month = aug,
       volume = {478},
       number = {3},
        pages = {3793-3808},
          doi = {10.1093/mnras/sty1231},
archivePrefix = {arXiv},
       eprint = {1805.04443},
 primaryClass = {astro-ph.GA},
       adsurl = {https://ui.adsabs.harvard.edu/abs/2018MNRAS.478.3793D}
}

@ARTICLE{Dobbs2014,
       author = {{Dobbs}, Clare and {Baba}, Junichi},
        title = "{Dawes Review 4: Spiral Structures in Disc Galaxies}",
      journal = {\pasa},
         year = 2014,
        month = sep,
       volume = {31},
          eid = {e035},
        pages = {e035},
          doi = {10.1017/pasa.2014.31},
archivePrefix = {arXiv},
       eprint = {1407.5062},
 primaryClass = {astro-ph.GA},
       adsurl = {https://ui.adsabs.harvard.edu/abs/2014PASA...31...35D}
}

@ARTICLE{Dobbs2008,
       author = {{Dobbs}, C.~L. and {Bonnell}, I.~A.},
        title = "{Simulations of spiral galaxies with an active potential: molecular cloud formation and gas dynamics}",
      journal = {\mnras},
         year = 2008,
        month = apr,
       volume = {385},
       number = {4},
        pages = {1893-1902},
          doi = {10.1111/j.1365-2966.2008.12995.x},
archivePrefix = {arXiv},
       eprint = {0801.3562},
 primaryClass = {astro-ph},
       adsurl = {https://ui.adsabs.harvard.edu/abs/2008MNRAS.385.1893D}
}

@ARTICLE{Block1994,
       author = {{Block}, D.~L. and {Bertin}, G. and {Stockton}, A. and {Grosbol}, P. and {Moorwood}, A.~F.~M. and {Peletier}, R.~F.},
        title = "{2.1 {\ensuremath{\mu}}m images of the evolved stellar disk and the morphological classification of spiral galaxies}",
      journal = {\aap},
         year = 1994,
        month = aug,
       volume = {288},
        pages = {365-382},
       adsurl = {https://ui.adsabs.harvard.edu/abs/1994A&A...288..365B}
}

@ARTICLE{Toomre1972,
       author = {{Toomre}, Alar and {Toomre}, Juri},
        title = "{Galactic Bridges and Tails}",
      journal = {\apj},
         year = 1972,
        month = dec,
       volume = {178},
        pages = {623-666},
          doi = {10.1086/151823},
       adsurl = {https://ui.adsabs.harvard.edu/abs/1972ApJ...178..623T}
}

@INPROCEEDINGS{Toomre1981,
       author = {{Toomre}, A.},
        title = "{What amplifies the spirals}",
    booktitle = {Structure and Evolution of Normal Galaxies},
         year = 1981,
       editor = {{Fall}, S.~M. and {Lynden-Bell}, D.},
        month = jan,
        pages = {111-136},
       adsurl = {https://ui.adsabs.harvard.edu/abs/1981seng.proc..111T}
}

@ARTICLE{Bland2016,
       author = {{Bland-Hawthorn}, Joss and {Gerhard}, Ortwin},
        title = "{The Galaxy in Context: Structural, Kinematic, and Integrated Properties}",
      journal = {\araa},
         year = 2016,
        month = sep,
       volume = {54},
        pages = {529-596},
          doi = {10.1146/annurev-astro-081915-023441},
archivePrefix = {arXiv},
       eprint = {1602.07702},
 primaryClass = {astro-ph.GA},
       adsurl = {https://ui.adsabs.harvard.edu/abs/2016ARA&A..54..529B}
}

@ARTICLE{Pedregosa2011,
       author = {{Pedregosa}, Fabian and {Varoquaux}, Ga{\"e}l and {Gramfort}, Alexandre and {Michel}, Vincent and {Thirion}, Bertrand and {Grisel}, Olivier and {Blondel}, Mathieu and {M{\"u}ller}, Andreas and {Nothman}, Joel and {Louppe}, Gilles and {Prettenhofer}, Peter and {Weiss}, Ron and {Dubourg}, Vincent and {Vanderplas}, Jake and {Passos}, Alexandre and {Cournapeau}, David and {Brucher}, Matthieu and {Perrot}, Matthieu and {Duchesnay}, {\'E}douard},
        title = "{Scikit-learn: Machine Learning in Python}",
      journal = {Journal of Machine Learning Research},
         year = 2011,
        month = oct,
       volume = {12},
        pages = {2825-2830},
          doi = {10.48550/arXiv.1201.0490},
archivePrefix = {arXiv},
       eprint = {1201.0490},
 primaryClass = {cs.LG},
       adsurl = {https://ui.adsabs.harvard.edu/abs/2011JMLR...12.2825P}
}

@ARTICLE{Kruskal1956,
    author = {Kruskal, Joseph B.},
    title = {On the shortest spanning subtree of a graph and the traveling salesman problem},
    journal = {Proceedings of the American Mathematical Society},
    volume = {7},
    number = {1},
    pages = {48--50},
    year = {1956},
    publisher = {American Mathematical Society}
}

\clearpage
\begin{appendix}
\nolinenumbers

\section{Analysis configuration and robustness checks}
\label{app:robustness}

This appendix presents the numerical configuration used in the baseline analysis and the sensitivity checks performed on the adopted BGMM and MST parameters. These tests are not intended as a full hyperparameter optimisation. Their purpose is to verify whether the main morphological interpretation depends on a finely tuned set of parameter values.

The input working samples contain 730 OC catalogue entries and 1120 OC+YSO entries. After the finite-value and robust filtering applied internally in the $(\theta_G,\ln R_G)$ plane, the BGMM fits use 729 and 1077 entries, respectively. The small difference between the input and fitted counts should therefore be understood as part of the preprocessing, rather than as an additional astrophysical selection.

The limitations of this baseline approach are clear. The analysis is morphological and mostly two-dimensional; it does not test age gradients, vertical coherence, or full kinematics. It also does not determine whether the observed segmentation is a formation signature of the spiral pattern or the subsequent outcome of shear, feedback, or transient spiral evolution. Addressing those questions requires information beyond the scope of the present work, including catalogue-dependent stability tests, velocities, vertical structure, gas counterparts, and comparisons with dynamical simulations.

The four pruning lengths in Table~\ref{tab:mst_sensitivity} are used only as a compact sensitivity grid. The behaviour is gradual rather than abrupt. In both samples, increasing $d_{\rm cut}$ increases the number of tracers belonging to retained branches and progressively merges smaller branches into larger, connected structures. This behaviour is particularly clear in the OC+YSO sample, where the largest branch grows from 135 nodes at $150\,{\rm pc}$ to 251 at $200\,{\rm pc}$, 325 at $250\,{\rm pc}$, and 694 at $300\,{\rm pc}$. These intermediate values document that the inferred connectivity changes progressively, rather than emerging only for a finely tuned pruning length.

\begin{table}[t]
\centering
\caption{Baseline configuration used for the BGMM segmentation and MST connectivity analysis.}
\label{tab:baseline_config}
\begin{tabular}{lp{4.0cm}}
\toprule
Quantity & Baseline choice \\
\midrule
\multicolumn{2}{l}{Coordinates and preprocessing} \\
Fitted coordinates & $(\theta_G,\ln R_G)$ \\
Angular unit in fits & radians \\
Feature scaling & standardised coordinates \\
Input filtering & age selection, robust $Z$ clipping, finite values, and robust Stage~1 filtering in $(\theta_G,\ln R_G)$ \\
\midrule
\multicolumn{2}{l}{KDE and BGMM segmentation} \\
KDE grid size & $250 \times 250$ \\
KDE smoothing & Gaussian smoothing, $\sigma=0.25$ \\
KDE peak quantile & $0.70$ \\
Initial BGMM components & KDE/BIC-guided candidate set \\
Covariance model & full covariance matrices \\
Random seed & 42 \\
BGMM relative-weight cut & $0.30$ \\
Additional relevance filters & effective population and density support \\
Protected components & significant KDE-peak components \\
\midrule
\multicolumn{2}{l}{Component membership} \\
Core Mahalanobis level & $\sigma_{P_1}=1.0$ \\
Envelope Mahalanobis level & $\sigma_{P_2}=2.25$ \\
Reference-arm labels & assigned only after BGMM segmentation \\
\midrule
\multicolumn{2}{l}{MST connectivity} \\
MST geometry & heliocentric $(X,Y)$ \\
MST selection & all finite tracer positions \\
MST pruning scales shown & $200$ and $300\,{\rm pc}$ \\
Minimum MST branch size & $N_{\rm min}=7$ nodes \\
\bottomrule
\end{tabular}
\end{table}

\begin{table}[t]
\centering
\caption{MST sensitivity to the pruning length.}
\label{tab:mst_sensitivity}
\begin{tabular}{lp{1cm}l p{1.5cm}p{1.25cm}}
\toprule
Sample & $d_{\rm cut}$ [pc] & Branches & Connected nodes & Largest branch \tabularnewline
\midrule
\multirow{4}{*}{OC} & 150 & 18 & 280 & 56 \tabularnewline
 & 200 & 20 & 416 & 75 \tabularnewline
 & 250 & 17 & 505 & 89 \tabularnewline
 & 300 & 9 & 596 & 187 \tabularnewline
\midrule
\multirow{4}{*}{OC+YSO} & 150 & 19 & 509 & 135 \tabularnewline
 & 200 & 18 & 704 & 251 \tabularnewline
 & 250 & 12 & 779 & 325 \tabularnewline
 & 300 & 9 & 864 & 694 \tabularnewline
\bottomrule
\end{tabular}
\tablefoot{For each sample and value of $d_{\rm cut}$, the table gives the number of retained branches, the number of tracers belonging to those branches, and the size of the largest branch. Branches are retained only if they contain at least $N_{\rm min}=7$ nodes.}
\end{table}

The BGMM sensitivity test is summarised in Table~\ref{tab:bgmm_sensitivity}. The OC sample retains seven components across all tested configurations. The OC+YSO sample yields nine components for the baseline, peak-threshold, and $P_2$ tests; this increases to ten components with a lower relative-weight cut, and decreases to seven with a stricter cut. The number of individual inter-arm points depends more heavily on the adopted $P_2$ envelope, as expected, since these points are defined within the outer regions of the components. The candidate intermediate Local--Sagittarius component is successfully recovered in both the baseline and the nearby peak-threshold and $P_2$ configurations, but it is pruned under the strictest relative-weight cut. Consequently, we interpret this feature as a candidate bridge or inter-arm structure supported by the baseline density field and the MST connectivity, rather than as a parameter-independent structure.

\begin{table*}[h]
\centering
\caption{BGMM sensitivity to nearby parameter choices.}
\label{tab:bgmm_sensitivity}
\begin{tabular}{llcccccc}
\toprule
Sample & Configuration & Peak q. & Ratio cut & $\sigma_{P_2}$ & Components & Inter-arm comp. & Inter-arm points \tabularnewline
\midrule
\multirow{7}{*}{OC} & baseline & 0.70 & 0.30 & 2.25 & 7 & 0 & 59 \tabularnewline
 & lower peak & 0.65 & 0.30 & 2.25 & 7 & 0 & 59 \tabularnewline
 & higher peak & 0.75 & 0.30 & 2.25 & 7 & 0 & 59 \tabularnewline
 & lower ratio & 0.70 & 0.25 & 2.25 & 7 & 0 & 59 \tabularnewline
 & higher ratio & 0.70 & 0.35 & 2.25 & 7 & 0 & 59 \tabularnewline
 & narrow $P_2$ & 0.70 & 0.30 & 2.00 & 7 & 0 & 18 \tabularnewline
 & wide $P_2$ & 0.70 & 0.30 & 2.50 & 7 & 0 & 93 \tabularnewline
\midrule
\multirow{7}{*}{OC+YSO} & baseline & 0.70 & 0.30 & 2.25 & 9 & 1 & 174 \tabularnewline
 & lower peak & 0.65 & 0.30 & 2.25 & 9 & 1 & 174 \tabularnewline
 & higher peak & 0.75 & 0.30 & 2.25 & 9 & 1 & 174 \tabularnewline
 & lower ratio & 0.70 & 0.25 & 2.25 & 10 & 1 & 181 \tabularnewline
 & higher ratio & 0.70 & 0.35 & 2.25 & 7 & 0 & 127 \tabularnewline
 & narrow $P_2$ & 0.70 & 0.30 & 2.00 & 9 & 1 & 128 \tabularnewline
 & wide $P_2$ & 0.70 & 0.30 & 2.50 & 9 & 1 & 238 \tabularnewline
\bottomrule
\end{tabular}
\tablefoot{The baseline configuration uses peak quantile $0.70$, relative-weight cut $0.30$, $\sigma_{P_1}=1.0$, and $\sigma_{P_2}=2.25$. The table reports the number of retained BGMM components, the number of whole components flagged as inter-arm candidates, and the number of individual points flagged as inter-arm or bridge-envelope points.}
\end{table*}

\end{appendix}

\end{document}